\documentclass[twocolumn,showpacs,preprintnumbers,amsmath,amssymb,superscriptaddress]{revtex4}

\usepackage{graphicx}
\usepackage{dcolumn}
\usepackage{bm}
\usepackage{graphics}
\usepackage{amsmath}
\usepackage{amssymb}
\usepackage{amscd}
\usepackage{afterpage}
\usepackage{float,times}
\usepackage{subfigure}
\usepackage{rotating}
\usepackage{multirow}
\usepackage{epsfig}
\usepackage{theorem}
\usepackage{moreverb}
\usepackage{euscript}
\usepackage{psfrag}

\begin{document}
\newcommand{\TeV}{\textrm{ TeV}}
\newcommand{\GeV}{\textrm{ GeV}}
\newcommand{\meV}{\textrm{ meV}}
\newcommand{\veps}{\varepsilon}
\newcommand{\lrarrow}{\leftrightarrow}
\newcommand{\order}{\mathcal{O}}
\def\slash{\not\!}
\newcommand{\ud}{\textrm{d}}
\title{Combining Left-Right And Quark-Lepton Symmetries In 5D}

\author{Andrew Coulthurst}
  \email[Email: ]{acoul@physics.unimelb.edu.au}
   \affiliation{School of Physics, Research Centre for High Energy Physics,\\
    University of Melbourne, Parkville, Victoria 3010, Australia.}
\author{Jason Doukas}
  \email[Email: ]{j.doukas@physics.unimelb.edu.au}
   \affiliation{School of Physics, Research Centre for High Energy Physics,\\
    University of Melbourne, Parkville, Victoria 3010, Australia.}
\author{Kristian L. McDonald}
  \email[Email: ]{klmcd@triumf.ca}
   \affiliation{School of Physics, Research Centre for High Energy Physics,\\
    University of Melbourne, Parkville, Victoria 3010, Australia.}
   \affiliation{Theory Group, TRIUMF, 4004 Wesbrook Mall, Vancouver, BC V6T2A3, Canada.}
\begin{abstract}
A five dimensional model containing both left-right and quark-lepton
symmetries is constructed, with the gauge group broken by a combination of
orbifold compactification and the Higgs
mechanism. An analysis of the gauge and scalar sectors is performed
and it is shown
that the 5d model admits a simpler scalar sector. Bounds on the
relevant symmetry breaking scales are
obtained and reveal that two
neutral gauge bosons may appear in the TeV energy range to be
explored by the LHC. Split
fermions are employed to remove the mass relations implied by the
quark-lepton symmetry and the necessary fermion localisation is achieved by
introducing bulk scalars with kink vacuum profiles. The
symmetries of the model constrain the Yukawa
sector, which in turn severely constrains the extent to which realistic
split fermion scenarios may be realized in the absence of Yukawa
coupling hierarchies. Nevertheless we present two interesting one
generation constructs. One of these provides a rationale for $m_t>m_b,
m_{\tau}$ and $m_\nu\ll m_t$ with Yukawa parameters which vary by only
a
factor of five. The other also suppresses the proton decay rate by
spatially separating quarks and leptons
but requires a Yukawa parameter hierarchy of order $10^2$.
 
\end{abstract}

\date{\today}

\pacs{11.10.Kk, 11.15.Ex,11.30Ly,14.60.Hi,14.70.Pw}
\maketitle


\section{Introduction\label{sec:orb_ql_intro_intro}}
Whilst the Standard Model (SM) of particle physics is a very
successful theory, it must be admitted that the
fermionic sector of the model contains a
rather strange set of gauge group representations. One of the
triumphs of the SM is the fact that this set of fermion
representations conspires to ensure the model is free from both
gauge and global anomalies. Though anomaly cancellation provides a
very strong motivation for the necessity of the observed
representations, it does not provide any insight into their
origin. Many questions remain, including `Why are left-chiral fermions
distinguished from right-chiral fermions?', `Why are quarks and
leptons different?' and `What is the origin of the strange set of
hypercharge values?'.

The Left-Right (LR)~\cite{Mohapatra:1986uf} and Quark-Lepton (QL)~\cite{Foot:1990dw} symmetric models
were introduced in an effort to answer these
questions; the hope being that the somewhat awkward arrangement of
fermions found in the SM may be simplified by a more fundamental
theory possessing a larger degree of symmetry. Both
the LR and QL models simplify the structure of the SM fermion representations
by relating previously disparate fields. However it is not until these
symmetries are combined together in the so called Quark-Lepton Left-Right
(QLLR) symmetric model that the fermionic sector of the SM simplifies
dramatically. Indeed with both QL and LR symmetries the assumed existence of
one SM field mandates the existence of all other SM fields from the
same generation~\cite{Foot:1989wt,Foot:1991fk}.

The goal of Grand
Unified Theories (GUTs) is to unite the SM forces into one larger
gauge group and reduce the number of independent fermion
representations. It is a remarkable feature of the QLLR model that the
quantum numbers of quarks and leptons may be unified independent of
gauge unification. This interesting fact makes it possible that
fermionic unification may be observed at low (TeV) energies even
if gauge unification does not occur until a very large energy scale.

A standard problem which arises when one extends the SM to obtain a
greater degree of symmetry is that the symmetry breaking sector of the
model must also be extended. This issue is often coupled with the
method of mass generation in the neutrino sector, with one stage of
high energy symmetry breaking assumed to result from a scalar field
which possesses the quantum numbers necessary to couple to a right-chiral
neutrino Majorana bilinear. Both the $SU(2)_R$ scalar triplet in
the LR model and the leptonic colour $[SU(3)_l]$ sextet scalar in
the QL model
serve the dual purpose of partially breaking the gauge symmetry and
generating a large right-chiral neutrino Majorana mass.

This has the
desirable consequence of allowing the see-saw mechanism to be implemented,
but at the expense of introducing scalars which do not
transform as a fundamental representation of the gauge
group. The problem becomes even more severe in the QLLR model;
one introduces seventy two additional complex scalar degrees of freedom
in the form of gauge representations which transform as chiral
$SU(2)$ triplets and colour $SU(3)$
sextets. The QLLR symmetry ensures that once one assumes the existence
of a scalar which induces a right-chiral neutrino Majorana mass three
additional scalar multiplets are also required.

The large interest in extra dimensional models in recent years has
uncovered new mechanisms to achieve symmetry breaking.
Using orbifold symmetry reduction allows one to reduce the bulk
symmetry to some subgroup operative at low energies (or the zero mode
level). This can reduce the number of scalars required in a model and
thus simplify the symmetry breaking sector. However the removal of
scalars (and in particular the reduction of the SM cutoff in models
which seek to solve the hierarchy problem) often removes the see-saw
mechanism as a viable source of neutrino mass suppression.

Fortunately the inclusion of additional spatial dimensions permits new
mechanisms for suppressing neutrino masses. In particular split
fermions allow one to suppress fermion masses relative to the
electroweak scale by spatially separating left- and right-chiral
fermions in an additional dimension~\cite{Arkani-Hamed:1999dc}. The subsequent reduction in higher
dimensional wave function overlap serves to suppress the effective Yukawa
coupling constants in the 4d theory.

In this work we investigate the implementation of these two mechanisms
in the context of a QLLR model. The objective of our paper is to
retain the attractive fermionic unification found in QLLR models but
reduce the complicated symmetry breaking sector required in 4d
constructs. We find that the scalar sector of the model may be
significantly reduced in 5d with the seventy degrees of freedom
previously mentioned not required to achieve a realistic low energy
model. The symmetry breaking sector of our model has the additional
consequence of permitting the two exotic neutral gauge bosons found in
QLLR models, $Z'$ and $Z''$, to appear at TeV energies and thus be
observable at the LHC. Previous works
have all required one of these bosons to be unobservably heavy.

The use of split fermions has a number of interesting
consequences. We provide two distinct one generational constructs
that suppress neutrino masses to experimentally
acceptable values and also provide a rationale for the inequalities $m_t>m_b,
m_{\tau}$. However the large degree of symmetry in the model
severely constrains the Yukawa sectors and it is a
non-trivial task to obtain fermion localisation patterns which account
for the range of fermion masses observed \emph{and} remove the need
for Yukawa parameter hierarchies.

We show that one generation of
flavour can be accounted for with Yukawa parameters which vary
only by a factor of five. However this setup does not allow one to
suppress proton decay by spatially separating quarks and leptons and
thus, along with the majority of split fermion works completed to date, the
model must be extended to avoid the usual hierarchy problem associated with
stabilizing the electroweak scale. In the alternative construct
the proton decay rate is safely suppressed by separating quarks and
leptons, but a Yukawa hierarchy of order $10^2$ is necessary to
achieve one generation of flavour. Thus one
may alleviate the hierarchy problem by lowering the cutoff to order
100~TeV. Further work is required to see if these promising results can
be carried
over to a full three generation model.

We note that recent works have investigated the LR
model~\cite{Mimura:2002te,Mohapatra:2002rn,Perez:2002wb} and the QL
model~\cite{McDonald:2006dy,Coulthurst:2006kz,Coulthurst:2006zu} in 5d.
The concept of leptonic colour has
also been generalised in~\cite{Foot:2006ie} and studied within the
context of unified theories
in~\cite{Joshi:1991yn,Babu:2003nw,Chen:2004jz,Demaria:2005gk,Demaria:2006uu,Demaria:2006bd}.

The layout of this paper is as follows. In
Section~\ref{sec:qllr_review} we review the main features of
the QLLR model. Section~\ref{sec:qllr_gauge_sector} details
the symmetry breaking sector of our 5d construct and
Section~\ref{sec:neutralgauge} looks at the gauge sector. The fermionic sector
is detailed in Section~\ref{sec:sf}, where we briefly
describe the features of split fermion models required for our
investigations and then present two promising one generation
fermionic geographies.  In
Section~\ref{qllr:neutral_currents} we discuss neutral currents and
derive bounds
on the symmetry breaking scales of the model. We consider some
experimental signatures of the model in Section~\ref{qllr:exp} and conclude in Section~\ref{qllr_conc}.
\section{Review Of The Quark-Lepton Left-Right Symmetric Model\label{sec:qllr_review}}
In this section we review the four dimensional QLLR
model~\cite{Foot:1989wt,Foot:1991fk}. To this end, let us recall
some features of the SM, the LR model and the QL model. The fermion spectrum of the SM
is given by:
\begin{eqnarray}\label{qllr_sm_fermions}
& &Q_L\sim (3,2,1/3),\mkern8mu u_R\sim (3,1,4/3),\mkern8mu
d_R\sim(3,1,-2/3),\nonumber\\
& &\\
& &L_L\sim (1,2,-1),\mkern8mu e_R\sim
(1,1,-2),\nonumber
\end{eqnarray}
where we have suppressed generational indices and the quantum numbers
label the transformation properties of the fields under
$G_{SM}\equiv SU(3)_c\otimes SU(2)_L\otimes U(1)_Y$. Whilst GUTs provide
us with a candidate explanation for the origin of the SM fermion
quantum numbers, it is safe to say that we do not yet know the
underlying theory responsible for the rather curious
collection of quantum numbers in (\ref{qllr_sm_fermions}). The key
observation made in GUTs is that the SM quantum numbers may be
understood if one embeds the group $G_{SM}$ into a simple group $H$.
The SM fermions are embedded into one (or two in the case of
$SU(5)$) representation $R$ of $H$. By employing
a suitable symmetry breaking mechanism to reduce the symmetry
operative at observable energy scales from the unifying group $H$
down to $G_{SM}$, the SM
fermion quantum numbers may be understood in terms of the
decomposition of $R$ under the low energy group $G_{SM}$.

An
alternative approach employed to uncover candidate extensions of the
SM follows from the observation that there exist suggestive similarities
amongst the quantum numbers of the SM fermions. One similarity is that
all left- and right-chiral fields possess identical electric and colour
charges; another is the similar family structure of quarks and leptons,
with all left-chiral fields forming $SU(2)_L$ doublets whilst their
right-chiral partners assume singlet $SU(2)_L$ representations. 

The motivation for the LR and QL models arises from these observed
similarities. By positing that the observed
similarity between left- and right-handed fields is the result of
an underlying symmetry one is lead to the LR
model~\cite{Mohapatra:1986uf}. This requires one to increase the fermion
content of the SM to include $\nu_{R}$ and to extend the gauge group
from $G_{SM}$ to $G_{LR}\equiv SU(3)_c\otimes SU(2)_L\otimes SU(2)_R
\otimes U(1)_{B-L}$. This extension has the desirable consequence of
simplifying the structure of the SM fermion content. The
fermion spectrum of the LR model is:
\begin{eqnarray}
& &Q_L\sim (3,2,1,1/3),\mkern20mu Q_R\sim(3,1,2,1/3),\nonumber\\
& &\\
& &L_L\sim (1,2,1,-1),\label{qllr_lr_fermions}\mkern20mu L_R\sim
(1,1,2,-1),\nonumber
\end{eqnarray} 
and the LR model Lagrangian is taken to be invariant under a discrete
$Z_2$ symmetry,
which we label as $Z_2^{LR}$, and whose action is defined by
\begin{eqnarray}
L_L\leftrightarrow L_R,\mkern10mu Q_L \leftrightarrow
Q_R,\mkern10muW_L\leftrightarrow W_R,
\end{eqnarray}
where $W_{L(R)}$ denotes the $SU(2)_{L(R)}$ gauge bosons. Observe that
the extended model reduces the total number of fermion
representations and also reduces the number of independent $U(1)$
charges per generation from five to two. The group $G_{LR}$ is broken
to ensure that $U(1)_Y\subset SU(2)_R\otimes U(1)_{B-L}$ so that the
SM is recovered at low energies.

If one instead focuses on the similar
family structures of quarks and leptons in the SM (assuming three $\nu_R$s) and follows the same
procedure one arrives at the
QL model~\cite{Foot:1990dw}. This requires $G_{SM}$ to be extended to $G_{QL}\equiv SU(3)_l\otimes SU(3)_c\otimes SU(2)_L
\otimes U(1)_X$ where $SU(3)_l$ is known as lepton colour and is the
leptonic equivalent of $SU(3)_c$ in the quark sector. As well as
adding $\nu_R$ to the SM fermion spectrum one must also triple the
number of leptons, giving the fermion spectrum
\begin{eqnarray}
& &Q_L\sim (1,3,2,1/3),\mkern20mu L_L\sim(3,1,2,-1/3),\nonumber\\
& &u_R\sim (1,3,1,4/3),\mkern20mu E_R\sim (3,1,1,-4/3),\label{qlllr_ql_fermion}\\
& &d_R\sim (1,3,1,-2/3),\mkern20mu N_R\sim (3,1,1,2/3).\nonumber
\end{eqnarray}  
The usual lepton $SU(2)_L$ doublet is contained in $L_L$ and $e_R$
($\nu_R$) is found inside $E_R$ ($N_R$). The Lagrangian of the QL
model permits a discrete symmetry, $Z_2^{QL}$, defined as follows:
\begin{eqnarray}
& &Q_L\leftrightarrow L_L\mkern10mu E_R\leftrightarrow u_R,\mkern10mu
N_R\leftrightarrow d_R,\nonumber\\
& &\\
& &G_c\leftrightarrow G_l, \mkern10mu C\leftrightarrow -C\nonumber
\end{eqnarray}
where $G_c$ ($G_l$) denotes the $SU(3)_{c(l)}$ gauge bosons and $C$ is
the $U(1)_X$ gauge boson. This model reduces the number of independent
$U(1)$ charges relative to the SM and also reduces the number of independent fermion representations
per generation from five (in the SM) to three. As with the LR model, the total
number of fermions per generation is greater than that of the SM due
to the exotics required to permit the defining discrete symmetry of
the model.

One may combine the symmetries $Z_2^{QL}$ and $Z_2^{LR}$ to
obtain the QLLR model. The gauge group of this model is
\begin{eqnarray}
G_{QLLR}= SU(3)_l\otimes SU(3)_c\otimes SU(2)_L\otimes SU(2)_R
\otimes U(1)_V\nonumber
\end{eqnarray}
and the fermions are assigned to the following representations:
\begin{eqnarray}
& &L_L\sim (3,1,2,1,-1/3),\mkern17mu L_R\sim(3,1,1,2,-1/3),\nonumber\\
& &\label{qllr_fermi_spectrum}\\
& &Q_L\sim(1,3,2,1,1/3),\mkern24mu Q_R\sim(1,3,1,2,1/3).\nonumber
\end{eqnarray}
The action of the discrete symmetry
$Z_2^{QL}\times Z_2^{LR}$ is defined as follows
\begin{equation}
\begin{array}{ccccccccccc}
L_L & \leftrightarrow & L_R &\qquad& V &\qquad& G_l &\qquad& W_L&\leftrightarrow&W_R\\
\updownarrow & ~&\updownarrow &\qquad& \updownarrow&\qquad &
\updownarrow&\qquad& ~& ~& ~\\
Q_L  & \leftrightarrow & Q_R &\qquad & -V  &\qquad&
G_q &\qquad& ~& ~& ~
\end{array}
\label{eqn:modelsymm}
\end{equation}
where the QL (LR) symmetry acts vertically (horizontally) and $V$
denotes the $U(1)_V$ boson. Note that (\ref{qllr_fermi_spectrum})
contains only one independent fermion field with the quantum numbers
of all other fermion fields determined completely by the discrete
symmetry. It is interesting that unification of the quark and lepton
quantum numbers may be achieved in the QLLR model, independent of
gauge coupling
unification. This is contrary
to the usual expectation that relationships which may exist between
the quark and
lepton quantum numbers are the manifestation of a symmetry which is
operative only at the GUT scale.

The simplified fermion content of the QLLR model comes at the expense
of an extended
scalar content. Both the $SU(3)_l$ and $SU(2)_R$ symmetries must be
broken to reproduce the SM at low energies. This breaking proceeds in
two steps. The first step is achieved by the introduction of the scalars
\begin{eqnarray}
& &\Delta_{1L}\sim (\bar{6},1,3,1,2/3),\mkern17mu
\Delta_{1R}\sim(\bar{6},1,1,3,2/3),\nonumber\\
& &\label{qllr_scalar_delta}\\
& &\Delta_{2L}\sim(1,\bar{6},3,1,-2/3),\mkern17mu \Delta_{2R}\sim(1,\bar{6},1,3,-2/3),\nonumber
\end{eqnarray}
which transform as
\begin{equation}
\begin{array}{ccc}
\Delta_{1L} & \leftrightarrow & \Delta_{1R}\\
\updownarrow & ~&\updownarrow \\
\Delta_{2L}  & \leftrightarrow & \Delta_{2R} 
\end{array}
\label{eqn:qllr_delta_scalars}
\end{equation}
under the discrete symmetries. The Yukawa Lagrangian for these fields
is
\begin{eqnarray}
\mathcal{L}_{\Delta}&=&\lambda_\Delta[\overline{(L_L)^c}L_L\Delta_{1L}+\overline{(L_R)^c}L_R\Delta_{1R}+
\overline{(Q_L)^c}Q_L\Delta_{2L}\nonumber\\
& &\qquad+\overline{(Q_R)^c}Q_R\Delta_{2R}] + \mathrm{H.c.},
\end{eqnarray}
Provided the neutral component of $\Delta_{1R}$ develops a non-zero
VEV the gauge symmetry will be broken as per
\begin{equation}
\begin{array}{c}
G_{QLLR}\\ \downarrow \\SU(2)_l\otimes SU(3)_c\otimes SU(2)_L\otimes U(1)_Y
\otimes U(1)_{Y'},
\end{array}\nonumber
\end{equation}
where $Y$ denotes the SM hypercharge, $Y'$ denotes some orthogonal
unbroken $U(1)$ factor whose precise form will not be important to
us and $SU(2)_l\subset SU(3)_l$. The hypercharge generator is
given by
\begin{equation}
 Y=2 I_{3R}+\frac{1}{\sqrt{3}} T_{l}^8 + V,\label{qllr_hypercharge_generator}
\end{equation}
where $T_{l}^8=1/\sqrt{3} \times \rm diag(-2,1,1)$ is a diagonal
generator of $SU(3)_l$ and $I_{3R}=1/2 \times \rm diag(1,-1)$ is
the diagonal generator of $SU(2)_R$.
Further symmetry breaking is accomplished by including the usual
colour triplet scalars found in QL models, namely
\begin{eqnarray}
\chi_l\sim(3,1,1,1,2/3),\mkern20mu \chi_q\sim(1,3,1,1,-2/3),\nonumber
\end{eqnarray}
which form partners under the $Z_2^{QL}$ symmetry, $\chi_l\leftrightarrow
\chi_q$. The Yukawa Lagrangian for these fields is
\begin{eqnarray}
\mathcal{L}_\chi&=&\lambda_\chi[\overline{(L_L)^c}L_L\chi_l+\overline{(L_R)^c}L_R\chi_l+
\overline{(Q_L)^c}Q_L\chi_q\nonumber\\
& &\qquad+\overline{(Q_R)^c}Q_R\chi_q] + \mathrm{H.c.}.
\end{eqnarray}
When the electrically neutral component of $\chi_l$ develops a VEV the following symmetry breaking occurs
\begin{equation}
\begin{array}{c}
SU(2)_l\otimes SU(3)_c\otimes SU(2)_L\otimes U(1)_Y
\otimes U(1)_{Y'}\\\downarrow\\ SU(2)_l\otimes SU(3)_c\otimes SU(2)_L\otimes U(1)_Y.
\end{array}\nonumber
\end{equation}
Note that $SU(2)_l$ remains unbroken. Whilst a large
number of additional scalars are required to achieve the desired
symmetry breaking it should be pointed out that only two additional
Yukawa couplings are introduced. The symmetries highly constrain the
Yukawa Lagrangian and, though we shall not need to consider it, they
also constrain the scalar
potential. Let us discuss briefly
the spectrum of exotic fermions and gauge bosons expected in the
model.

The VEV hierarchy $\langle
\Delta_{1R}\rangle\gg\langle\chi_l\rangle$ is assumed as the
nonzero value for $\langle \Delta_{1R}\rangle$ induces
a Majorana mass for the right-handed neutrinos. After neutrinos
acquire a Dirac mass at the
electroweak symmetry breaking scale the seesaw mechanism will thus be operative
to suppress the observed neutrino masses below the electroweak
scale. The nonzero VEV for $\langle \Delta_{1R}\rangle$ also gives mass to
the $SU(3)_l/SU(2)_l$ coset gauge bosons and the $W_R$
bosons. As the seesaw
mechanism requires $\langle
\Delta_{1R}\rangle$ to
be large, roughly $10^{14}$~GeV, these gauge bosons become
unobservably heavy. The VEV for $\Delta_{1R}$ also breaks the linear
combination of $I_{3R}$, $T_{l}$ and $V$ which is orthogonal to $Y$
and $Y'$. Thus a neutral boson $Z''$ gains a mass of
order $\langle\Delta_{1R}\rangle$. 

The non zero VEV for $\chi_l$ breaks $Y'$, resulting in a
massive neutral gauge boson with an order
$\langle\chi_l\rangle=w_l$ mass. The symmetry breaking induced
by $\chi_l$ also gives mass to the exotic fermions introduced to
fill out the $SU(3)_l$ fermion representations. These fermions are
known as liptons in the literature and are a common feature of models
possessing a QL symmetry. The unbroken $SU(2)_l$ symmetry serves to
confine the liptons into two-fermion bound states. These states all
decay via the usual electroweak interactions into the known fermions~\cite{Foot:1991fk}. The lower bound on $w_l$ is of order TeV
(we provide a detailed discussion of the bound on $w_l$
in Section~\ref{qllr:neutral_currents}) and the key experimental signatures for the
model are the $Z'$ boson and the liptons. The liptons may be produced
at the LHC via the usual electroweak interactions and via virtual $Z'$
creation. 

The gauge group $G_{SM}\otimes SU(2)_l$ must be broken down to
$SU(3)_c\otimes U(1)_Q\otimes SU(2)_l$. This requires the
introduction of a Higgs bidoublet
\begin{eqnarray}
\Phi\sim(1,1,2,2,0),
\end{eqnarray}
resulting in the following electroweak Yukawa Lagrangian
\begin{eqnarray}
\mathcal{L}_\Phi&=&\lambda_{\Phi1}[\bar{L}_LL_R\Phi +
\bar{Q}_LQ_R\tilde{\Phi}]+\nonumber\\
& &\lambda_{\Phi2}[\bar{L}_LL_R\tilde{\Phi}+\bar{Q}_LQ_R\Phi]
+\mathrm{H.c.},\label{eq:phi_yukawa_lagrangian}
\end{eqnarray}
where $\tilde{\Phi}=\epsilon \Phi^* \epsilon$ (we denote the
two dimensional anti-symmetric tensor as $\epsilon$) and
\begin{eqnarray}
\Phi\leftrightarrow\tilde{\Phi}
\end{eqnarray}
under the QL symmetry. If the neutral components of $\Phi$ develop a
VEV the desired symmetry breaking is achieved. The Yukawa couplings in
(\ref{eq:phi_yukawa_lagrangian}) give rise to fermion Dirac masses and
result in mass relations of the type
\begin{eqnarray}
m_u=m_e,\qquad m_d=m_\nu,\label{qllr_4d_mass_relations}
\end{eqnarray}
where $m_\nu$ is the neutrino Dirac mass. As the light neutrinos
acquire mass via the seesaw mechanism the relationship $m_d=m_\nu$
doesn't provide any phenomenological difficulty. The right-handed
neutrinos acquire a Majorana mass through their couplings to
$\Delta_{1R}$ and there is enough parameter freedom in the Lagrangian
$\mathcal{L}_{\Delta}$ to ensure that arbitrary neutrino mass values
can be obtained. The relationship between the down quark mass matrix
and the neutrino Dirac mass matrix actually serves to reduce the
number of parameters employed to implement the seesaw mechanism. The mass relations between the electrons and the up
quarks may also be removed by introducing an additional
bidoublet $\Phi'\sim(1,1,2,2,0)$. This doubles the number of
Yukawa couplings and thus also nullifies the mass relations $m_d=m_\nu$,
thereby reducing predictivity of the model.
\section{Symmetry Breaking In Five Dimensions\label{sec:qllr_gauge_sector}}
In this work we study the quark-lepton left-right symmetric
extension to the Standard Model in five dimensions. The additional
spatial dimension is
taken as the orbifold $S^1/Z_2\times Z_2'$, whose coordinate is
labelled as $y$. The construction of the orbifold proceeds via the
identification $y\rightarrow -y$ under the $Z_2$ symmetry and
$y'\rightarrow-y'$ under the $Z_2'$ symmetry, where $y'=y+\pi
R/2$. The physical region in $y$ is given by the interval $[0,\pi
R/2]$. 

Given the absence
of chirality in five dimensions we shall denote the
gauge group of the theory as $SU(3)_l\times SU(3)_c\times
SU(2)_1 \times SU(2)_2 \times U(1)_V$. We will be required
to ensure that the low energy fermion spectrum contains the chiral
fermions found in the SM. The
zero mode $SU(2)_{1(2)}$ gauge bosons will eventually be identified with the usual $W_{L(R)}$ bosons
in LR models via their action on the low energy fermion content. Thus
the 5d theory is invariant under the interchange $1\leftrightarrow2$
which will prove to be equivalent to the usual LR symmetry in the low
energy theory. This
matter has already been discussed in~\cite{Mohapatra:2002rn}. We shall continue to label the discrete symmetry of the
5d model as $Z_2^{QL}\times Z_2^{LR}$.

The orbifold action also has a definition on the space of gauge fields
which propagate in the bulk. We define $P$ and $P'$ to be matrix 
representations of the orbifold actions $Z_2$ and $Z_2'$ respectively. To maintain gauge invariance under these projections, 
the gauge fields must have the transformations
\begin{eqnarray}
A_{\mu}(x^{\mu},y) & \rightarrow & A_{\mu}(x^{\mu},-y) = P A_{\mu}(x^{\mu},y) P^{-1},\\ \nonumber
A_{5}(x^{\mu},y) & \rightarrow & A_{5}(x^{\mu},-y) = - P A_{5}(x^{\mu},y) P^{-1}, \\ \nonumber
A_{\mu}(x^{\mu},y') & \rightarrow & A_{\mu}(x^{\mu},-y') = P' A_{\mu}(x^{\mu},y') P'^{-1}, \\ \nonumber
A_{5}(x^{\mu},y') & \rightarrow & A_{5}(x^{\mu},-y') = - P'
A_{5}(x^{\mu},y') P'^{-1},\label{eq:qllr_projections}
\end{eqnarray}
where $A$ denotes the bulk gauge sector,
\begin{widetext}
\begin{eqnarray}
A^M(x^{\mu},y)&=&G^M_l (x^{\mu},y)\oplus G^M_q(x^{\mu},y)\oplus
W^M_1(x^{\mu},y)\oplus W^M_2(x^{\mu},y)\oplus V^M(x^\mu,y), \nonumber\\
&=& G_l^{M\,a} \,T^a \oplus G_q^{M\,a}\, T^a\oplus
W_1^{M\,i}\, \tau^i\oplus W_2^{M\,i}\tau^i\oplus V, 
\end{eqnarray}
\end{widetext}
with $M$ being the 5d Lorentz index, $T$ denotes the $SU(3)$ generators, $\tau$ denotes the $SU(2)$
generators and the
gauge indices take the values $a=1,2,...,8$ and $i=1,2,3$. 

Given that $P$ and $P'$ define a representation of reflection symmetries their eigenvalues are $\pm 1$. We can 
express these matrices in diagonal form, with a freedom in the parity choice of the entries. 
The exact nature of these actions then completely determines the gauge
symmetry which remains unbroken in the low energy limit of the theory
(namely the zero mode gauge sector). 
Unless $P$ is the identity matrix, not all the gauge fields will commute 
with the orbifold action. 
These fields will not possess a zero mode, and thus only a subset of the 
5d gauge theory is manifest at the zero mode level. 
Ideally, the bulk gauge group would reduce to $G_{SM}\otimes
SU(2)_l$ at the zero mode level; however, this is not directly possible via orbifolding. 
The $Z_2 \times Z_2'$ actions are abelian and commute with the
diagonal gauge group generators. 
Subsequently, the rank of the bulk gauge group must be conserved at
the zero mode level. This means that breaking unwanted 
$SU(3)$ and $SU(2)$ factors has the trade-off of retaining spurious $U(1)$
subgroups and one must invoke a mechanism in tandem to 
orbifolding in order to accomplish the breaking to $G_{SM}\otimes SU(2)_l$. 

The orbifold action can be decomposed as:
\begin{eqnarray}
& &(P,P^\prime)=\nonumber\\
& &(P_l\oplus P_q
\oplus P_1
\oplus P_2 \oplus P_V,~ P^\prime_l\oplus P'_q \oplus P'_1 \oplus
P'_2\oplus P'_V),\nonumber
\end{eqnarray}
with
\begin{eqnarray}
P_l&=&\text{diag}(1,1,1), ~ P_l'=\text{diag}(-1,1,1),\nonumber\\
P_q&=&P_q'= \text{diag}(1,1,1),\nonumber\\
P_1&=&P_1'= \text{diag}(1,1),\nonumber\\
P_2&=& \text{diag}(1,1),~P_2'=\text{diag}(-1,1)\nonumber\\
P_V&=&P_V'=\rm 1.\label{eq:gauge_parities}
\end{eqnarray}
The only non-trivial entries occur in the $SU(2)_2$ and $SU(3)_l$ 
gauge space. Denoting these gauge fields as
\begin{equation}\label{eqn:Wmatrix}
W_2 = \frac{1}{2}\left(
\begin{array}{cc}
W_2^0 & \sqrt2 W_2^+ \\
\sqrt2 W_2^- & -W_2^0
\end{array}
\right),
\end{equation}
and
\begin{equation}\label{eqn:Gmatrix}
G_l= \left(
\begin{array}{ccc}
-\frac{2}{\sqrt{3}} G_l^{0} & \sqrt{2} Y_l^1 & \sqrt{2}Y_l^2\\
\sqrt{2}Y_l^{1\dag} & G_l^3+\frac{1}{\sqrt{3}}G_l^{0} & \sqrt{2}\emph{\~{G}}_l \\
\sqrt{2} Y_l^{2\dag} & \sqrt{2} \emph{\~{G}}_l^{\dag} & -G_l^{3}
+\frac{1}{\sqrt{3}} G_l^{0}
\end{array}\right),
\end{equation}
the $Z_2 \times Z_2^\prime$ parities of these fields is found to be
\begin{eqnarray}
G^0_{l\mu},~ G^3_{l\mu},~ \emph{\~{G}}_{l\mu},~ \emph{\~G}^{\dag}_{l\mu},~ W^0_{2\mu} :~ (+,+), \label{Eqn:zeromodes}\\
Y^1_{l\mu},~ Y^2_{l\mu},~ Y^{1\dag}_{l\mu},~ Y^{2\dag}_{l\mu},~ W^\pm_{2\mu} :~ (+,-), \label{Eqn:exoticbosons1}\\
G^0_{l5},~ G^3_{l5}, ~\emph{\~{G}}_{l5},~ \emph{\~G}^{\dag}_{l5}, ~W^0_{2,5} :~ (-,-), \label{Eqn:exoticbosons2}\\
Y^1_{l5},~ Y^2_{l5},~ Y^{1\dag}_{l5},~ Y^{2\dag}_{l5},~
W^\pm_{2,5} :~ (+,-).\label{Eqn:exoticbosons3}
\end{eqnarray}
A general five dimensional field, $\psi$, can
be expanded in terms of Fourier modes in the compact dimension:
\begin{eqnarray}\nonumber
\psi_{(+,+)}(x^\mu, y) &=& \sqrt{\frac2{\pi R}}\psi_{(+,+)}(x^\mu)+\nonumber\\& & \frac{2}{\sqrt{\pi R}} \sum_{n=1}^{\infty}
\psi_{(+,+)}^{(n)}(x^\mu) \cos \frac{2ny}R,\nonumber\\ \nonumber
\psi_{(+,-)}(x^\mu, y) &=& \frac2{\sqrt{\pi R}}
\sum_{n=1}^{\infty} \psi_{(+,-)}^{(n)}(x^\mu) \cos
\frac{(2n-1)y}R,\\ \nonumber
\psi_{(-,+)}(x^\mu, y) &=&
\frac2{\sqrt{\pi R}} \sum_{n=1}^{\infty} \psi_{(-,+)}^{(n)}(x^\mu) \sin \frac{(2n-1)y}R, \\
\psi_{(-,-)}(x^\mu, y) &=& \frac2{\sqrt{\pi R}}
\sum_{n=1}^{\infty} \psi_{(-,-)}^{(n)}(x^\mu) \sin
\frac{2ny}R.\nonumber
\end{eqnarray}
Thus only fields with a $(+,+)$ parity under $Z_2
\times Z_2^\prime$ posses a massless zero mode. Importantly,
we see that $G^0_\mu,~ G^3_\mu,~ \emph{\~{G}}_\mu,~
\emph{\~G}^{\dag}_{\mu},~ W^0_\mu$ are the only such four dimensional gauge
fields to do so. Effectively then our $SU(3)_l\times SU(2)_2$
symmetry has been broken down to $SU(2)_l\times U(1)_l\times
U(1)_2$ at the zero mode level. It is worth commenting that new
heavy exotic bosons, corresponding to fields with $(+,-)$, $(-,+)$ and
$(-,-)$ parities, exist in
Kaluza Klein (KK) states at the inverse compactification scale, along with
the KK towers for the fields with $(+,+)$ parities. The complete zero
mode gauge group is thus
\begin{equation}\nonumber
SU(2)_l\times SU(3)_c\times SU(2)_1 \times U(1)_l\times U(1)_2
\times U(1)_V.
\end{equation}
The parities in (\ref{eq:gauge_parities}) ensure that all
fifth dimensional components of the bulk gauge fields do not possess
a zero mode. Consequently no spurious scalars appear in the low energy
theory.

The remaining symmetry breaking shall be achieved via the Higgs
mechanism. This requires the following scalars
\begin{eqnarray}
& &\chi_l\sim(3,1,1,1,2/3),\mkern15mu\chi_q\sim(1,3,1,1,-2/3),\nonumber\\
& &\chi_1\sim(1,1,2,1,1),\qquad \chi_2\sim(1,1,1,2,1),\nonumber\\
& &\Phi\sim(1,1,2,2,0),
\end{eqnarray}
which we take to be bulk fields. As in the 4d case $\chi_l\leftrightarrow\chi_q$ under
$Z_2^{QL}$ and $\chi_1\leftrightarrow\chi_2$ under $Z_2^{LR}$. We do
not define the transformations of the Higgs bidoublet under
$Z_2^{QL}\times Z_2^{LR}$ at this stage. Under $Z_2 \times Z_2^\prime$ we assume the Higgs fields transform
as:
\begin{eqnarray*}
\chi_2 (x^\mu, y) &\rightarrow& \chi_2 (x^\mu, -y) = P_2 \chi_2 (x^\mu,y), \\
\chi_2 (x^\mu, y^\prime) &\rightarrow& \chi_2 (x^\mu, -y^\prime) = P_2^\prime \chi_2 (x^\mu,y^\prime), \\
\chi_1 (x^\mu, y) &\rightarrow& \chi_1(x^\mu, -y) = P_1 \chi_1 (x^\mu,y), \\
\chi_1 (x^\mu, y^\prime) &\rightarrow& \chi_1(x^\mu, -y^\prime) = -P_1^\prime \chi_1 (x^\mu,y^\prime), \\
\chi_q (x^\mu, y) &\rightarrow& \chi_q(x^\mu, -y) = P_q \chi_q (x^\mu,y), \\
\chi_q (x^\mu, y^\prime) &\rightarrow& \chi_q(x^\mu, -y^\prime) = P_q^\prime \chi_q (x^\mu,y^\prime), \\
\chi_l (x^\mu, y) &\rightarrow& \chi_l(x^\mu, -y) = P_l \chi_l (x^\mu,y), \\
\chi_l (x^\mu, y^\prime) &\rightarrow& \chi_l(x^\mu, -y^\prime) = -P_l^\prime \chi_l (x^\mu,y^\prime), \\
\Phi(x^\mu,y)&\rightarrow& \Phi(x^\mu,-y)=P_1\Phi (x^\mu,y)P_2^{-1},\\
\Phi(x^\mu,y')&\rightarrow& \Phi(x^\mu,-y')=P_1'\Phi (x^\mu,y)P_2'^{-1},
\end{eqnarray*}
where the matrix representations of the orbifold reflection symmetries
in the scalar sector are necessarily the same as those introduced for
the gauge sector in (\ref{eq:gauge_parities}). The parity assignments for the bulk scalar
fields immediately follow:
\begin{equation}
\chi_{l}=\left(
\begin{array}{c}
\chi_{1l}^0(+,+)\\
\chi_{2l}^{+1/2}(+,-)\\
\chi_{3l}^{+1/2}(+,-)\\
\end{array}\right)
,\quad\chi_{2}=\left(
\begin{array}{c}
\chi_{1,2}^+(+,-)\\
\chi_{2,2}^0(+,+)
\end{array}\right),
\end{equation}
\begin{equation}
\Phi= \left(
\begin{array}{cc}
                                      \phi_{1}^0(+,+) & \phi_{2}^-(+,-) \\
                                       \phi_{1}^+(+,+)  &
                                       \phi_{2}^0(+,-)
                                      \end{array} \right).
\end{equation}
We denote the VEVs of the zero mode scalars as
\begin{equation}
\langle \chi_{l}^{(0)} \rangle =\left(
\begin{array}{c}
w_l\\
0\\
0\\
\end{array} \right),\
\langle \chi_2^{(0)} \rangle = \left(
\begin{array}{c} 0 \\ w_R
\end{array} \right), \ \langle \Phi^{(0)} \rangle = \left(
\begin{array}{cc}
                                      k & 0 \\
                                       0  & 0
                                      \end{array} \right).
\label{Higgs vevs}
\end{equation}
The subscript $R$ on the $\chi_{2}$ VEV has been used to adopt the
familiar four dimensional notation. Observe that we have not included
the four $\Delta$ scalars
in (\ref{qllr_scalar_delta}). These seventy-two degrees of freedom
have been replaced with the four degrees of freedom contained in
$\chi_{1,2}$. All scalars in the 5d model form fundamental
representations of the
gauge group. This has the advantage of decoupling the $SU(3)_l$ and
the $SU(2)_2$ symmetry breaking scales. The nonzero value for $w_l$
induces the breaking:
\begin{eqnarray}
G_{QLLR}\rightarrow SU(2)_l\otimes G_{LR},
\end{eqnarray}
whilst the VEV for $\chi_2$ gives:
\begin{eqnarray}
G_{QLLR}\rightarrow G_{QL}.
\end{eqnarray}
Thus in the limit $w_l\rightarrow\infty$ ($w_R\rightarrow\infty$) we
essentially reproduce the usual 5d orbifold broken LR (QL) model. Note that both $w_l$
and $w_R$ may be of order TeV (as we shall discuss further in
Section~\ref{qllr:neutral_currents}), which will provide one of the distinctions between our
construct and 4d QLLR models studied to date. Previous models have
required the scalars $\Delta$ to permit the seesaw mechanism to be
operative. As we shall see in Section~\ref{sec:sf} the higher
dimensional theory permits an alternative mechanism for suppressing
neutrino masses relative to the electroweak scale. Thus we may
consider the model without the additional degrees of freedom required
to implement the seesaw mechanism.
\section{The Gauge Sector\label{sec:neutralgauge}}
In this section we discuss the
phenomenology of the gauge sector in detail. Let us first consider the
charged bosons. We shall henceforth identify the $SU(2)$ bosons in
terms of their action on the zero mode fermion spectrum, i.e.
$W_1\rightarrow W_L$ and $W_2\rightarrow W_R$. The charged
gauge bosons do not mix and have the KK mass towers:
\begin{eqnarray}
m^2_{n,W^\pm_L}&=&\frac{g^2}{2}k^2+\left(\frac{2n}{R}\right)^2,\\
m^2_{n,W^\pm_R}&=&\frac{g^2}{2}(k^2+w_R^2)+\left(\frac{2n+1}{R}\right)^2,\\
m^2_{n,Y^1}&=&\frac{g_s^2}{2}w_l^2+\left(\frac{2n+1}{R}\right)^2,\\
m^2_{n,Y^2}&=&\frac{g_s^2}{2}w_l^2+\left(\frac{2n+1}{R}\right)^2,
\end{eqnarray}
where $n=0,1,2...$. The mass of the
lightest $Y^1$, $Y^2$ and $W_R$ bosons
are set by the inverse compactification scale and only $W_L$ has
a zero mode. We shall work under the assumption that $w_{l,R}\ll
1/R$ and thus the only light charged boson is $W_L^{\pm(0)}$, with
all other charged bosons first appearing at energies of order $1/R$.

The
neutral gauge bosons do mix with each other and we denote their mass terms as:
\begin{equation}
\mathcal{L}_\textrm{mass}=\frac{1}{2} \sum_n V\mathcal{M}_n^2
V^{\dagger}, \end{equation}
where $V=(W_L^{0(n)},W_R^{0(n)},
B_V^{(n)}, G_l^{0(n)})$, and
\begin{widetext}
\begin{equation}
\mathcal{M}_n^2=\left(
\begin{array}{cccc}
\frac{g^2 k^2}{2} +(\frac{2n}{R})^2& -\frac{g^2 k^2}{2}       & 0                                   & 0\\
-\frac{g^2 k^2}{2}& \frac{g^2 (k^2+w_R^2)}{2} +(\frac{2n}{R})^2& -\frac{g_v g w_R^2}{2}               & 0\\
0                 &-\frac{g_v g w_R^2}{2} &
g_v^2(\frac{w_R^2}{2}+\frac{2w_l^2}{9}) +(\frac{2n}{R})^2&
-\frac{2g_v g_s
w_l^2}{3\sqrt{3}}\\
0                 &0                         &-\frac{2g_v g_s
w_l^2}{3\sqrt{3}}& \frac{2g_s^2 w_l^2}{3}+(\frac{2n}{R})^2\\
\end{array}\right).
\end{equation}
\end{widetext}
Only the zero mode gauge bosons possess masses less than $1/R$ and
under our hierarchy $1/R\gg w_{l,R}$ we may neglect the higher
modes. In order to simplify the analysis it is useful to introduce the SM
$U(1)_Y$ field with coupling constant $g_Y$ and a $U(1)_{B-L}$
field with coupling $g_B$. In
these terms the coupling constants are
related by,
\begin{eqnarray}\nonumber
\frac{1}{e^2}=\frac{1}{g^2}+\frac{1}{g_Y^2},\\ \nonumber
\frac{1}{g_Y^2}=\frac{1}{g^2}+\frac{1}{g_B^2},\\
\frac{1}{g_B^2}=\frac{1}{g_V^2}+\frac{1}{3g_s^2},\label{Eqn:couplings}
\end{eqnarray}
and the fields
\begin{eqnarray}\label{eqn:A}
A&=&\cos\theta~ B_Y+\sin\theta~ W^0_{L},\nonumber\\
Z&=&-\sin\theta~ B_Y+\cos\theta~
W^0_{L},\nonumber\\
~\nonumber\\
B_Y&=&\cos\alpha ~ B_B+\sin\alpha ~W^0_{R},\nonumber\\
Z'&=&-\sin\alpha ~ B_B+\cos\alpha ~
W^0_{R},\nonumber\\
~\nonumber\\
B_B&=&\cos\beta~ B_V+\sin\beta~ G_l^0,\nonumber\\
Z''&=&-\sin\beta ~B_V+\cos\beta ~G^0_l \label{eqn:zdoubleprime},
\end{eqnarray}
where the mixing angles are defined as
\begin{eqnarray}\nonumber
\textrm{tan}~\theta=\frac{g_Y}{g}, \quad \textrm{tan} ~\alpha
=\frac{g_B}{g} \quad \textrm{and} \quad \textrm{tan}~
\beta=\frac{g_V}{\sqrt{3} g_s}.\\
\label{Eqn:angles} ~
\end{eqnarray}
Using Equations (\ref{Eqn:angles}) and (\ref{Eqn:couplings}) one
can relate all angles to the Weinberg angle, $\theta$.
\begin{eqnarray}
\sin\alpha &=&\tan\theta,\\
\sin\beta &=&
\frac{g}{\sqrt{3}g_s}\frac{\sin\theta}{\sqrt{\cos2\theta}}.\label{eqn:beta}
\end{eqnarray}
Expressing the zero mode neutral boson masses in terms of the fields
(\ref{eqn:zdoubleprime}) reveals a massless photon ($A$) and mixing
between the remaining bosons. Writing $\vec{Z}=(Z, Z', Z'')^T$ and $
\mathcal{L}_\textrm{mass}=\frac{1}{2}\vec{Z}^{\dagger} H \vec{Z}$ one has
\begin{widetext}
\begin{equation}\label{eqn:ZMixing}
H=\left(
\begin{array}{ccc}
m_Z^2 & -m_Z^2\cot\alpha ~\sin\theta & 0\\
-m_Z^2\cot\alpha ~\sin\theta & m_{Z'}^2 & \frac{g^2w_R^2}{4}\tan 2\theta \tan\beta\\
0  &\frac{g^2w_R^2}{4}\tan 2\theta \tan\beta & m^2_{Z''}
\end{array}\right),
\end{equation}
\end{widetext}
where,
\begin{eqnarray}
m_Z^2 &=&\frac{g^2 k^2}{2\cos^2\theta},\nonumber\\
m_{Z'}^2 &=& \frac{g^2w_R^2}{2
\cos^2\alpha}\left(1+\left(\frac{k}{w_R}\right)^2\cos^4\alpha \right),\nonumber\\
m_{Z''}^2 &=&
\frac{2g_s^2}{3\cos^2\beta}\left(w_l^2+\frac{9}{4}w_R^2\sin^4\beta
\right).\label{eqn:MassesLargewl}
\end{eqnarray}
Letting $w$ generically denote $w_l$ and $w_R$, the physical mass-squared eigenvalues
to $\mathcal{O}\left(\frac{k^2}{w^2}\right)$ are:
 \begin{eqnarray}
M_{Z}^2 &=& m_Z^2,\\
M_{Z^{\prime}}^2 &=& M^2-\frac{1}{2}\Delta-m_Z^2 \mu_{+} \cos^2\alpha\cos^2\theta,\\
M_{Z^{\prime\prime}}^2 &=& M^2+\frac{1}{2}\Delta+m_Z^2
\mu_{-} \cos^2\alpha\cos^2\theta,
\end{eqnarray}
with
\begin{eqnarray}
M^2&\equiv&A w_l^2+ Bw_R^2,
\end{eqnarray}
and
\begin{equation}
\frac{1}{2}\Delta=\sqrt{A^2{w_l}^4 +
     C{w_l}^2{w_R}^2  +
     B^2{w_R}^4},
\end{equation}
where
\begin{eqnarray}
A &=&\frac{1}{3} {g_s}^2{\sec^2\beta},\\
B &=&\frac{{1}}{4}\,\left( g^2\,{\sec^2\alpha } + 3\,{
{g_s}}^2\,{\sin^2\beta}\,{\tan^2\beta } \right),\\
C &=&A(2B-g^2\sec^2\alpha) ,
\end{eqnarray}
and
\begin{eqnarray}
\mu_{\pm}&=&\frac{1}{2}\pm\frac{1}{4}\left(3g_s^2\sin^2\beta\tan^2\beta-g^2\sec^2\alpha\right)\frac{w_R^2}{\Delta}\nonumber\\
& &\pm\frac{1}{3}g_s^2\sec^2\beta\frac{w_l^2}{\Delta}. 
\end{eqnarray}
The leading order correction to the $Z$ mass is obtained by retaining
higher order terms, giving
\begin{eqnarray}
M_Z^2= m_Z^2 +\delta m _Z^2,
\end{eqnarray}
where
\begin{eqnarray}\label{eqn:zmasscorrection}
\delta
m_Z^2&=&-m^2_Z\left[\left(\frac{k}{w_R}\right)^2\cos^4\alpha+\frac{g^4}{4g_s^4}\left(\frac{k}{w_l}\right)^2\tan^4\theta\right].\nonumber\\
& &
\end{eqnarray}
It is unnecessary to determine the higher order corrections to the
$Z'$ and $Z''$ masses. The physical $Z$-bosons are found by performing
a 3-dimensional rotation of
the interaction $Z$-bosons:
\begin{eqnarray}\label{eqn:urotation_text}
\left(\begin{array}{c} Z_{phy}\\
Z_{phy}'\\Z_{phy}''\end{array}\right)=U^{-1}\left(\begin{array}{c}
   Z\\Z'\\Z''\end{array}\right).
\end{eqnarray}
We present the $3\times 3$ mixing matrix $U$ in
Appendix~\ref{appendix_a}. Using
the results
from Appendix~\ref{appendix_a} one may verify the usual LR and
QL behaviour of the neutral gauge sector in the various large $w$
limits. For completeness we note that in the large $w_l$ limit we
find
\begin{eqnarray}
M_{Z}^2 &=&m_Z^2 -\delta_{LR},\\
M_{Z^{\prime}}^2
&=&\frac{g^2w_R^2}{2\cos^2\alpha}\left[1+\left(\frac{k}{w_R}\right)^2\cos^4\alpha\right]
+\delta_{LR},\\
M_{Z^{\prime\prime}}^2
&=&\frac{2g_s^2w_l^2}{3\cos^2\beta},
\end{eqnarray}
where $\delta_{LR}=\left(\frac{k}{w_R}\right)^2\cos^4\alpha$. As expected, $Z'$ is the usual LR boson and
$M_{Z'}^2$ agrees
with~\cite{Mohapatra:2002rn} aside from a misprint in that work~\cite{mohapatra_error}. Furthermore, in this limit
$(U_{33})^2\rightarrow 1$
which implies that
\begin{eqnarray}
U\rightarrow\left(
\begin{array}{ccc}
 \cos \xi &
   \sin \xi &
   0  \\
   -\sin \xi &
   \cos \xi & 0 \\
   0 &
   0  & 1
   \end{array}\right),
   \end{eqnarray}
where we have defined $\xi=-(\zeta+\psi)$ in terms of the angles
(\ref{eqn:zeta}) and (\ref{eqn:psi}). We find
\begin{eqnarray}
\tan\xi &=& U_{21}/U_{11},\\
&=&\left(\frac{k}{w_R}\right)^2\frac{\sin\alpha\cos^3\alpha}{\sin\theta}.
\end{eqnarray}
in agreement with~\cite{Mohapatra:2002rn}.

In the large $w_R$ limit the heaviest physical neutral boson is a
linear combination of $Z'$ and $Z''$
(see Appendix~\ref{large_w_l_limit_appendix}). The mass eigenvalues are
\begin{eqnarray}
M_{Z}^2 &=&m_Z^2(1 -\frac{g^4}{4g_s^4}\left(\frac{k}{w_l}\right)^2\tan^4\theta),\\
M_{Z^{\prime}}^2
&=&\frac{\frac{2}{3}g_s^2w_l^2}{1-\tan^2\theta\frac{g^2}{3g_s^2}},\\
M_{Z^{\prime\prime}}^2
&=&\frac{1}{2}(g^2\sec^2\alpha+3g_s^2\sin^2\beta\tan^2\beta)w_R^2.
\end{eqnarray}
The mixing angle
between $Z$ and $Z'$ is given by (see
Appendix~\ref{large_w_l_limit_appendix})
\begin{equation}
\tan \zeta =
\frac{\sqrt{3}}{4}\left(\frac{k}{w_l}\right)^2\left(\frac{g}{g_s}\right)^3\frac{\tan^2\theta}{\cos\theta},
\end{equation}
a result which has not previously appeared in the literature.

\section{5d QLLR with Split Fermions}\label{sec:sf}
Having discussed in some detail the symmetry breaking and gauge
sectors of the model
we now turn our attention to the fermions. One
interesting aspect of studying models in additional dimensions is the
novel new mechanisms which become available to solve old problems. As
we have already emphasised, the scalar content of 4d QLLR models is
quite complicated with the $\Delta$ scalars of
equation (\ref{qllr_scalar_delta}) included to simultaneously break
the gauge symmetry and suppress neutrino masses below the electroweak
scale (via the seesaw mechanism). These states have not been included in the
5d construct and thus we must present an alternative method of
suppressing neutrino masses if we are to persist with the simplified
scalar content.  In doing this we will find we are also able to remove the troublesome mass relations which occur in 4d QLLR models without the need for a second Higgs bidoublet.

Since all
the fermions transform non-trivially under either $SU(3)_c$
or $SU(3)_l$ and either $SU(2)_1$ or $SU(2)_2$, their $Z_2\times
Z_2'$ transformations are given by:
\begin{eqnarray}
\Psi(x^\mu,y)\rightarrow \Psi(x^\mu,-y)=\pm\gamma_5P^a_{1,2}P^\alpha_{q,l}\Psi_{a,\alpha}(x^\mu,y),\nonumber \\
\Psi(x^\mu,y')\rightarrow
\Psi(x^\mu,-y')=\pm\gamma_5P'^a_{1,2}P'^\alpha_{q,l}\Psi_{a,\alpha}(x^\mu,y'),
\end{eqnarray}
where $a$ ($\alpha$) are indices of the relevant $SU(2)$ ($SU(3)$)
group. The $\pm$ signs in the two equations are independent and
govern which chiral component of the fermion wavefunction will be
odd and which even about the relevant fixed point. These orbifold boundary conditions (OBCs)
force the two $SU(2)_L$ quark/lepton singlets of the SM to
come from different $SU(2)_2$ doublets.  We must therefore
double the minimal fermion content of our model compared with 4d
QLLR models. This doubling of the fermion spectrum is typically
required in 5d
LR~\cite{Mohapatra:2002rn}
and QL models~\cite{Coulthurst:2006kz}. Thus the fermion spectrum is:
\begin{eqnarray}
& &L_1,L_1'\sim (3,1,2,1,-1/3),\nonumber\\
& &L_2,L_2'\sim(3,1,1,2,-1/3),\nonumber\\
& &Q_1,Q_1'\sim(1,3,2,1,1/3),\nonumber\\
& &Q_2,Q_2'\sim(1,3,1,2,1/3),\label{qllr_5d_fermi_spectrum}
\end{eqnarray}
where generation indices have been suppressed. The
symmetries of the QLLR model, together with the requirement that the
low energy spectrum match that of the SM (up to possible additional neutrinos),
strongly restrict the fermion orbifold parities. Preservation of the
$Q\leftrightarrow L$ and $1\leftrightarrow 2$
symmetries in the Lagrangian together with the zero mode content
requirements completely specifies the OBCs of the fermions as:
\begin{widetext}
\begin{equation}
\begin{array}{cc}
Q^{r,b,g}_{1,L}\sim\left(\begin{array}{c}u(+,+)\\d(+,+)\end{array}\right)^{r,b,g},\quad
&
Q'^{r,b,g}_{1,L}\sim\left(\begin{array}{c}u(+,-)\\d(+,-)\end{array}\right)^{r,b,g},\\
Q^{r,b,g}_{1,R}\sim\left(\begin{array}{c}u(-,-)\\d(-,-)\end{array}\right)^{r,b,g},\quad
&
Q'^{r,b,g}_{1,R}\sim\left(\begin{array}{c}u(-,+)\\d(-,+)\end{array}\right)^{r,b,g},\\
Q^{r,b,g}_{2,L}\sim\left(\begin{array}{c}u(-,+)\\d(-,-)\end{array}\right)^{r,b,g},\quad
&
Q'^{r,b,g}_{2,L}\sim\left(\begin{array}{c}u(-,-)\\d(-,+)\end{array}\right)^{r,b,g},\\
Q^{r,b,g}_{2,R}\sim\left(\begin{array}{c}u(+,-)\\d(+,+)\end{array}\right)^{r,b,g},
\quad&
Q'^{r,b,g}_{2,R}\sim\left(\begin{array}{c}u(+,+)\\d(+,-)\end{array}\right)^{r,b,g},
\end{array}\nonumber
\end{equation}
\begin{equation}
\begin{array}{cccc}
L^{r'}_{1,L}\sim\left(\begin{array}{c}\nu(+,-)\\e(+,-)\end{array}\right)^{r'},\quad&
L'^{r'}_{1,L}\sim\left(\begin{array}{c}\nu(+,+)\\e(+,+)\end{array}\right)^{r'},\quad&
L^{b',g'}_{1,L}\sim\left(\begin{array}{c}\nu(+,+)\\e(+,+)\end{array}\right)^{b',g'},\quad&
L'^{b',g'}_{1,L}\sim\left(\begin{array}{c}\nu(+,-)\\e(+,-)\end{array}\right)^{b',g'},\\

L^{r'}_{1,R}\sim\left(\begin{array}{c}\nu(-,+)\\e(-,+)\end{array}\right)^{r'},\quad&
L'^{r'}_{1,R}\sim\left(\begin{array}{c}\nu(-,-)\\e(-,-)\end{array}\right)^{r'},\quad&
L^{b',g'}_{1,R}\sim\left(\begin{array}{c}\nu(-,-)\\e(-,-)\end{array}\right)^{b',g'},\quad&
L'^{b',g'}_{1,R}\sim\left(\begin{array}{c}\nu(-,+)\\e(-,+)\end{array}\right)^{b',g'},\\

L^{r'}_{2,L}\sim\left(\begin{array}{c}\nu(-,+)\\e(-,-)\end{array}\right)^{r'},\quad&
L'^{r'}_{2,L}\sim\left(\begin{array}{c}\nu(-,-)\\e(-,+)\end{array}\right)^{r'},\quad&
L^{b',g'}_{2,L}\sim\left(\begin{array}{c}\nu(-,-)\\e(-,+)\end{array}\right)^{b',g'},\quad&
L'^{b',g'}_{2,L}\sim\left(\begin{array}{c}\nu(-,+)\\e(-,-)\end{array}\right)^{b',g'},\\

L^{r'}_{2,R}\sim\left(\begin{array}{c}\nu(+,-)\\e(+,+)\end{array}\right)^{r'},\quad&
L'^{r'}_{2,R}\sim\left(\begin{array}{c}\nu(+,+)\\e(+,-)\end{array}\right)^{r'},\quad&
L^{b',g'}_{2,R}\sim\left(\begin{array}{c}\nu(+,+)\\e(+,-)\end{array}\right)^{b',g'},\quad&
L'^{b',g'}_{2,R}\sim\left(\begin{array}{c}\nu(+,-)\\e(+,+)\end{array}\right)^{b',g'}.

\end{array}
\label{eqn:fermionorbifoldparities}
\end{equation}
\end{widetext}
Here the numerical subscripts and the primes are used to label
different 5d fields
such that $Q_{1L}$ and $Q_{1R}$ ($Q_{1L}'$ and $Q_{1R}'$) form the left and right chiral
components of the one 5d field $Q_1$ ($Q_1'$) etc., the superscripts
$r,b,g$ label quark colours and $r',b',g'$ label lepton colours. We
have taken $r'$ to be the colour of the SM leptons. Note that zero
modes of some of the
exotic $b'$ and $g'$ coloured leptons are present. The appearance of
these states
is a fortunate consequence of the fermion orbifold
parity structure as they are required to
ensure an anomaly free zero mode fermion content~\cite{Arkani-Hamed:2001is}.
These states gain masses as in the 4d theory via the $\chi$ Yukawa Lagrangian which has the form (we must define the action of the discrete symmetries before we can specify it exactly):
\begin{equation}
\mathcal{L}_{\textrm{Yuk}_\textrm{non-EW}}\sim
\sum_\textrm{fermions}h_i\left(\overline{L^c_i}L_i\chi_l+\overline{Q^c_i}Q_i\chi_q\right),\label{qllr_eq_non_ew_Yuk}
\end{equation}
whilst the quarks remain massless
since $\chi_q$ has a vanishing VEV. 
We must also define the action of the QL and
$1\leftrightarrow 2$ symmetries on the fermions.  Due to the doubling of the
fermion spectrum there are several ways we could do this.  The various
possibilities result in different phenomenology and influence the
extent to which the issues of neutrino mass, proton decay and unwanted
mass relations can be resolved. Below we shall investigate the two
most interesting scenarios.

We structure the remainder of this section as follows. In
Section~\ref{sub:split_fermion_intro} we briefly introduce
split
fermions. Split fermion models \cite{Arkani-Hamed:1999dc} use an inherently
extra dimensional construct to motivate the masses of SM
fermions and/or proton longevity. As we shall be employing split fermions to
address these issues a brief introduction is in order. In Section~\ref{sec:nmasspdecay} we discuss the nature of neutrino mass and proton decay in our model. In
Section~\ref{sec:sfmasses} we explore the use of split fermions
with one possible assignment for
quarks, leptons and scalars under the QL and $1\leftrightarrow2$ symmetries.
This assignment is interesting as it induces a fermion localisation
pattern motivating the differences in quark and lepton masses
observed in the SM. The hierarchy between, for example, the top quark
and a Dirac
neutrino is obtained with Yukawa couplings which vary by only a factor
of five.

We investigate an alternative symmetry assignment in Section~\ref{sec:sfpdecayandmasses}. This arrangement allows
one to simultaneously suppress the proton decay rate and understand the range
of fermion masses in the SM. We note that
the symmetries of the model highly constrain the parameters required
to localise fermions. It is a non-trivial result that we
are able to remove the unwanted mass relations implied by the QL
symmetry, suppress the proton decay rate by spatially separating
quarks and leptons and understand some of the flavour features found
in the SM. To the best of our knowledge this is the first model in the
literature that motivates a localisation pattern
which simultaneously ensures proton longevity and addresses flavour. We
demonstrate our ideas in this section with one generation
examples and further work is required to ensure that these promising
ideas carry over to a three generational model. A complete
numerical analysis of the three generational setup is beyond the scope
of the present work.
\subsection{Split Fermion Mechanism}\label{sub:split_fermion_intro}
In extra dimensional models the effective 4d theory is obtained by
integrating out the extra dimensions and any symmetry or naturalness
arguments should be made in the fundamental extra dimensional model.
The basic idea behind split fermion models is that by appropriately
choosing the profiles of the fields in the extra dimensions, the
conclusions of any symmetry or naturalness argument in the
extra dimensional model need not apply in the effective theory. In
their original paper, Arkani-Hamed and Schmaltz (AS)
\cite{Arkani-Hamed:1999dc} noted two situations where this observation
is useful: to explain the hierarchy in SM fermion masses and to
explain the
stability of the proton.

The work of AS was performed with an infinite extra dimension. We are
interested in the case where the extra dimension is compactified and therefore
follow \cite{Grossman:2002pb}.  To localise
fermions along the extra dimension we introduce a gauge singlet bulk scalar,
$\Sigma_1$, assumed to possess odd parity about both fixed points.

For a given bulk fermion, $\psi$, the Yukawa Lagrangian with the bulk
scalar $\Sigma_1$ is
\begin{equation}
\mathcal{L}=\overline\psi(i\gamma^M\partial_M-f_{\psi_1}\Sigma_1)\psi+\frac{1}{2}\partial^M\Sigma_1 \partial_M\Sigma_1-\frac{\Lambda_1}{4}(\Sigma_1^2-v_1^2)^2,
\end{equation}
where $f_{\psi_1}$, $\Lambda_1$ and $v_1$ are constants. The OBCs prevent $\Sigma_1$
from developing a constant VEV along the extra dimension and lead to a
kink configuration
\begin{equation}
\langle\Sigma_1\rangle\approx v_1 \tanh\left(\xi_1 v_1y\right)\tanh\left(\xi_1v_1\left(\frac{\pi R}{2}-y\right)\right),
\label{eqn:kink}
\end{equation}
where $\xi_1^2\equiv \Lambda_1/2$. Solving the Dirac Equation for the fermion gives:
\begin{equation}
\psi(y)=N e^{f_\psi\int^y_0\langle\Sigma_1\rangle(y')\ud y'}.
\end{equation}
Using \eqref{eqn:kink} this solution is approximately a Gaussian of
width $(f_{\psi_1}v_1)^{-1}$ localized around $y=0$ ($y=\pi R/2$) for $f_{\psi_1}
v_1>0$ ($f_{\psi_1} v_1<0$).  Thus by assuming distinct
couplings to $\Sigma_1$ for distinct SM fermion multiplets one can localize them around different fixed
points with varying widths.

The fermion $\psi$ may be shifted from the fixed points by
using two localizing scalars, $\Sigma_{1,2}$, with VEVs $v_{1,2}$ and
fermion couplings $f_{\psi_{1,2}}$. One finds that
$\psi$ is localized around $y=0$ ($y=\pi R/2$) for
$f_{\psi_1} v_1,f_{\psi_2} v_2>0$ ($f_{\psi_1} v_1,f_{\psi_2} v_2<0$).
However, if $\textrm{sign}(f_{\psi_1} v_1)\neq\textrm{sign}(f_{\psi_2}
  v_2)$, the localization of $\psi$ will depend on the relative sizes
of $f_{\psi_i} v_i$. Cases exist where the fermion is
localized around one of the fixed points, within the bulk or
has a bimodal profile.  Fermions localized inside the bulk generally
have wider profiles than those localized at a fixed point. A detailed
discussion of the various cases
may be found in~\cite{Grossman:2002pb}.

Having demonstrated the localization of fermions, we now discuss the
motivation of AS for doing so. To simplify the explanation we shall
assume that the fermion profiles are
exactly Gaussians of width $\mu^{-1}$. 

AS had two motivations for
localising fermions.
Firstly, since the left- and right-handed components of a given SM
fermion are in different gauge multiplets they can be localized at
different points in the extra dimension. The Higgs Yukawa coupling in
the effective 4d theory is
\begin{equation}
\mathcal{L}=f \int_0^{L} \ud y \overline{F_R} F_L \Phi=fkK\overline{f_R}f_L,
\end{equation}
where $L\equiv\pi R/2$ is the length of the extra dimension, $k$ is the Higgs VEV, $K=\int_0^{L}F_R(y) F_L(y)\ud y\sim
e^{-\mu^2r^2}$ and $r$ is the separation between the left and right
handed fields.  Thus even if the fundamental Yukawa coupling, $f$, is
of order one, that in the effective theory can be exponentially
suppressed. Since $\mu$ and $r$ will vary for different fermions
it is natural to expect the observed hierarchy in SM fermion
masses. Previous studies have
confirmed that it is possible to obtain the SM masses and mixings from
this setup with reduced parameter
hierarchies~\cite{Grossman:2002pb,SFFlavour}. The price we pay is that we must
introduce a new free parameter for every scalar-fermion coupling.  The setup therefore lacks
predictivity, telling us nothing, for example, about the relative
masses of the quarks and leptons, or the top and bottom quarks.

Secondly, the AS proposal allows one to
consider fundamental theories which contain non-renormalizable proton decay inducing
operators without insisting that the fundamental scale be very
large. Instead the operators may be suppressed in the effective theory
by localizing quarks and leptons at opposite ends of the extra
dimension. It was shown by AS that regardless of the particular proton
decay inducing operator, this leads to suppression of the rate of
proton decay going like $\sim e^{-\mu^2 {L^2}}$.  Thus
provided $\mu\gtrsim 10/L$ the proton lifetime will be greater
than the experimental lower bounds.  Whilst providing a novel
alternative explanation for the stability of the
proton, this requires that we arbitrarily choose
$\textrm{sign}f_{q_i}=-\textrm{sign}f_{l_j}$ for all $i,j$ where $i$
($j$) runs over all left- and right-handed SM quarks (leptons).

In calculating the localization of the fermions in our model below, we shall do so only classically.  In taking this approach we are following previous work on split fermions \cite{SFFlavour} in assuming that any quantum corrections will be small enough not to alter the qualitative nature of our fermion geography. Before attempting to find realistic split fermion geographies for our model, we briefly discuss the nature of neutrino mass and proton decay within it. 
\subsection{Neutrino mass and proton decay}
\label{sec:nmasspdecay}
The nature of neutrino mass and proton decay in any model are related to the status of baryon and lepton number symmetries. Our model contains accidental unbroken baryon- and lepton-number symmetries. Baryon number, $B$, takes the value $1/3$ for quarks and $-2/3$ for the coloured scalar $\chi_q$. Lepton number is given by
\begin{equation}
L=L'-\frac{T^8_l}{\sqrt3},
\end{equation}
where $L'$ is 1/3 for leptons and $-2/3$ for $\chi_l$.  It is easy to check that all renormalizable Lagrangian terms which respect the gauge symmetries also conserve $B$ and $L$.  As such there is no process which will lead to Majorana neutrino mass or proton decay at any order within the model. This differs from 4d QLLR models where the larger scalar sector precludes lepton number conservation.

However, there are non-renormalizable operators which may induce these processes.  The leading order effective operators resulting in Majorana neutrino masses are
\begin{align}
\order^\textrm{eff}_{\nu_L}&=\frac{g}{(\Lambda\pi R)^3}\frac{(\chi_l^\dagger\chi_R\tilde\Phi L_L)^2}{\Lambda^5},\nonumber\\
\order^\textrm{eff}_{\nu_R}&=\frac{g'}{(\Lambda\pi R)^2}\frac{(\chi_lN_R^c\chi_R^\dagger)^2}{\Lambda^3}.
\end{align}
These operators can only lead to cosmologically acceptable masses for the right-handed neutrino while keeping the left-handed neutrinos light if we take the breaking scales to be at least $w_{l,R}\gtrsim100\TeV$.  For phenomenological reasons, such a possibility is uninteresting so we do not consider it further.  Instead we will assume that these Majorana masses are zero or negligibly small.  This could occur either if the cut-off, $\Lambda$, is large, or if a sub-group of the $B$ and $L$ symmetries is preserved also above the cut-off so as to forbid these operators.

Proton decay occurs non-renormalizably via the effective operator
\begin{equation}
\order^\textrm{eff}_p=\frac{1}{(\Lambda\pi R)^{3/2}}\frac{\epsilon_{\alpha\beta\gamma}Q^\alpha Q^\beta Q^\gamma \chi_{l\alpha'}^\dagger L^{\alpha'}}{\Lambda^{3}}.\label{eqn:QLLRpdecay}
\end{equation}
Below we will consider two possibilities for preventing proton decay.  In Section \ref{sec:sfmasses} we will assume this operator is unimportant either because the cut-off is high, or the high energy theory respects the $B$ symmetry evident at low energies.  In Section \ref{sec:sfpdecayandmasses}, we consider the case where this operator can lead to significant proton decay and must be suppressed via the split fermion mechanism \cite{BLsymm}.
\subsection{Fermion mass relationships}\label{sec:sfmasses}
In the original 4d QLLR models with a single Higgs bidoublet the QL
symmetry led to phenomenologically inconsistent mass relations between
the quarks and leptons.  Depending on how we define the action of the
discrete symmetries on the fermions, these can be partially removed in
our 5d model due to the doubling of the fermion spectrum. To proceed
any further we must define the action of the $Q\lrarrow L$ and
$1\lrarrow 2$ on the fermions, which we take to be
\begin{equation}
\begin{array}{ccccccc}
L_{1} & \leftrightarrow & L_{2} &\qquad& L'_{1} & \leftrightarrow & L'_{2}\\
\updownarrow & ~&\updownarrow &\qquad& \updownarrow & ~&\updownarrow\\
Q_{1}  & \leftrightarrow & Q_{2} &\qquad & Q'_{1}  &
\leftrightarrow & Q'_{2}.
\end{array}
\label{eqn:fermionsymm1}
\end{equation}
If we take the Higgs bidoublet to transform trivially under $1\lrarrow
2$ and as $\Phi\lrarrow\tilde\Phi$ under QL, the resulting EW Yukawa
Lagrangian is
\begin{widetext}
\begin{equation}
\mathcal{L}_{\textrm{Yuk}_\textrm{EW}}=\lambda_{1} (\overline Q_1
\tilde\Phi Q_2 + \overline L_1 \Phi L_2) +\lambda_{2} (\overline
Q_1 \Phi Q_2' + \overline L_1 \tilde\Phi L_2') +\lambda_{3}
(\overline Q_1' \Phi Q_2 + \overline L_1' \tilde\Phi L_2)
+\lambda_{4} (\overline Q_1' \tilde\Phi Q_2' + \overline L_1'
\Phi L_2')+\textrm{H.c.}.
\end{equation}
\end{widetext}
Note that the $1\leftrightarrow
2$ symmetry requires
\begin{equation}
\lambda_{1}=\lambda_{1}^\dagger,\quad
\lambda_{4}=\lambda_{4}^\dagger,\quad
\lambda_{2}=\lambda_{3}^\dagger.
\label{eqn:yukrelations}
\end{equation}
The EW Yukawa Lagrangian for the SM particles,
\mbox{$\mathcal{L}_{\textrm{Yuk}_\textrm{EW}}^{\textrm{SM}}\subset\mathcal{L}_{\textrm{Yuk}_\textrm{EW}}$}, is
\begin{eqnarray}
\mathcal{L}_{\textrm{Yuk}_\textrm{EW}}^{\textrm{SM}}&=&\lambda_{1}k^*\overline
d_L d_R + \lambda_{2}k\overline u_L u_R+\lambda_{3}k^*\overline
e_L e_R \nonumber\\
& &+ \lambda_{4}k\overline \nu_L \nu_R +\textrm{H.c.}.
\end{eqnarray}
Thus Equation \eqref{eqn:yukrelations} implies $m_u=m_e$. As in the 4d case, these phenomenologically
incorrect relationships can be removed by introducing a second Higgs
bidoublet but at the cost of predictivity and without any
explanation for the hierarchical nature of SM fermion masses. 

Split fermions provide a natural alternative approach. Naively one may
think that the symmetries of our model over constrain the extra
dimensional fermion profiles.  Indeed with only one localizing
scalar, the quark doublet and the right-handed down quark and charged
lepton of a given generation necessarily have the same
profile, albeit possibly localized around different fixed points (and
similarly for the lepton doublet and right-handed up
quark and neutrino).
However, by using two localizing scalars with different parities under
the symmetries it is possible to give the fermions different
profiles and also move some fermions into the bulk. 

Taking $\Sigma_1$ to be even under both the $Q\leftrightarrow L$ and
$1\leftrightarrow 2$ symmetries and $\Sigma_2$ even (odd) under
$Q\leftrightarrow L$ ($1\leftrightarrow 2$) results in the Lagrangian
\begin{align}
\mathcal{L_{\textrm{Yuk}_{\textrm{kink}}}}&=
f(\overline{Q_1^{c}}Q_1
+\overline{L^{c}_1}L_1+\overline{Q_2^{c}}Q_2+\overline{L^{c}_2}
L_2)\Sigma_1\nonumber\\
&+f'(-\overline{Q'^{c}_1}Q'_1 -\overline{L'^{c}_1}L'_1-\overline{Q'^{c}_2}Q'_2-\overline{L'^{c}_2}L'_2)\Sigma_1\nonumber\\
&+g(\overline{Q_1^{c}}Q_1 +\overline{L^{c}_1}L_1-\overline{Q_2^{c}}Q_2-\overline{L^{c}_2}L_2)\Sigma_2\nonumber\\
&+g'(-\overline{Q'^{c}_1}Q'_1 -\overline{L'^{c}_1}L'_1+\overline{Q'^{c}_2} Q'_2+\overline{L'^{c}_2}L'_2)\Sigma_2.
\end{align}
Taking $f, f', g, g'>0,$ the coupling of the SM quark doublets
to both scalars is positive, strongly localizing them around the $y=0$
fixed point while the lepton doublets couple negatively ensuring they
are localized at the $y=L$ fixed point.  The SM singlets couple
to the two scalars with different signs allowing them to be localized
at either fixed point or within the bulk.  However the symmetries
ensure that the profiles of the right-handed up quarks and neutrinos
(down quarks and charged leptons) have identical extra dimensional
profiles. 

We note that the localisation of fermions along the extra dimension
does not suppress the mass of zero mode exotic
leptons
relative to $w_l$. Inspection of
eq. (\ref{qllr_eq_non_ew_Yuk}) reveals that the mass terms generated
by the $\chi_l$ VEV couple exotic leptons from the same gauge
multiplet. As these fields necessarily have the same profile in the
extra dimension the lightest exotic leptons are generically expected
to have an order $w_l$ mass independent of the localization
pattern required to achieve a realistic SM spectrum. 

\begin{figure}
\centering
\includegraphics[width=0.4\textwidth]{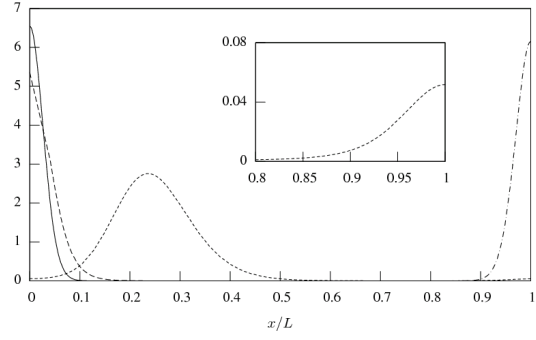}
  \caption{Fermion profiles for the parameter values described in Section
\ref{sec:sfmasses}. Solid line = $Q_L$, short dash = $b_R,\tau_R$, long
dash = $t_R,\nu_R$ and the dot-dash = $L_L$. Note that some of the fermions
possess identical profiles. In particular the inset shows the
identical profiles of
$b_R$ and $\tau_R$ about $x/L=1$.}
  \label{fig:splitfermions_masses}
\end{figure}

In order to give a concrete example, we consider a single generation
model. To determine the localisation pattern of fermions it is
only necessary to specify the bulk scalar parameters $v_{1,2}$ and
the Yukawa coupling
constants $f,f',g,g'$ as functions of $\xi_{1,2}$ and $L=\pi R/2$. We
take these to be $v_1=4/(\xi_1 L)$ and
$v_2=12/(\xi_2 L)$.  With the choice of fermion-$\Sigma$ couplings $f=28.4\xi_1,\quad f'=14.4\xi_1,
\quad g=7.0\xi_2\quad g'=6.4\xi_2, $. The resulting fermion
localization pattern is shown in
Figure \ref{fig:splitfermions_masses} and is of interest as it allows us to
explain several SM
features:
\begin{itemize}
        \item The top singlet is localized on top of the quark doublet so we
expect $m_t\approx k$, while the bottom singlet is in the bulk leading us
to expect $m_t>m_b$.  Since $b_R$ is localized in the bulk it has a
relatively large width. This ensures
that the suppression of $m_b$ is not too large.
        \item The tau singlet is localized in the bulk close to the opposite
fixed point to the lepton doublet leading to $m_t>m_\tau$.  Again the
large width of $\tau_R$ prevents the suppression being too large.
        \item The right handed neutrino is strongly localized with $t_R$
about
the opposite fixed point to the lepton doublet.  The strong localization
of both $\nu_L$ and $\nu_R$ allows the neutrino mass to be tiny.
\end{itemize}
Recalling that our
symmetries force the Higgs Yukawa coupling of the top and tau to be identical
(Eq. \eqref{eqn:yukrelations}\ ), the Yukawa couplings
$\lambda_t=\lambda_\tau=\lambda_b=\lambda_\nu=1.01$
lead to masses~\cite{running_masses} $m_t=169\GeV,\quad m_b=4.16\GeV,\quad m_\tau=1.77\GeV,\quad
m_\nu=26\meV$. Hence we are able to obtain realistic fermion masses
with fewer free parameters than previously required. Whilst a complete
three generation study remains to be undertaken, this approach does
appear to provide a viable and novel approach to explaining the SM fermion
masses with fewer free parameters and without any parameter
hierarchies. It also nullifies the phenomenologically incorrect mass
relationships of previous QLLR models.

\subsection{Simultaneously Suppressing Proton Decay and Obtaining Correct Fermion Masses}\label{sec:sfpdecayandmasses}
We have shown that our 5d QLLR setup enables us to obtain realistic
fermion masses. However this setup does not allow one to suppress the
proton decay rate. Proton decay occurs in the 5d QLLR model from operators
of the form
\begin{eqnarray}
O_p\sim\frac{1}{\Lambda^{9/2}}Q^3L\chi_l^\dagger,
\end{eqnarray}
where $Q$ ($L$) denotes a quark (lepton) field and $\Lambda$ is the
fundamental scale. As quarks and leptons have significant fifth
dimensional wavefunction overlap  with
the setup in
Section~\ref{sec:sfmasses} one must take the fundamental scale to be
large or extend the model to ensure proton longevity. If one simply assumes the cutoff is large the usual
fine tuning is required to stabilize the Higgs mass at the electroweak
scale.

It was shown
in~\cite{Coulthurst:2006bc} that models
with a QLLR symmetry admit a split fermion setup which suppresses
proton decay less arbitrarily than the split fermion implementation of
the SM. This requires one of the localizing scalars
to be odd (even) under Q~$\lrarrow$~L ($1\lrarrow 2$). Unfortunately neither of the scalars in
Section~\ref{sec:sfmasses} transformed in this way. If the
fermion transformations of
Eq. (\ref{eqn:fermionsymm1}) are retained and one of
the localizing scalars of Section~\ref{sec:sfmasses} is forced to be
odd (even) under the Q~$\lrarrow$~L ($1\lrarrow 2$) symmetry, the
resulting
fermionic geographies require large parameter hierarchies to produce
realistic mass spectra. We instead choose $\Phi$ to be trivial under QL and the fermions
to transform as
\begin{equation}
\begin{array}{ccccccc}
L_{1} & \leftrightarrow & L_{2} &\qquad& L'_{1} & \leftrightarrow & L'_{2}\\
\updownarrow & ~&\updownarrow &\qquad& \updownarrow & ~&\updownarrow \\
Q_{1}  & \leftrightarrow & Q'_{2} &\qquad & Q'_{1}  &
\leftrightarrow & Q_{2}
\end{array}
\label{eqn:fermionsymm2}
\end{equation}
under the QL and $1\leftrightarrow2$ symmetries which leads to the
mass relationship
$m_d=m_e$. Choosing $\Sigma_1$ ($\Sigma_2$) to be odd (odd) under the
$Q\lrarrow L$ symmetry and even (odd) under the $1\lrarrow 2$
symmetry, the localizing scalar Yukawa Lagrangian is
\begin{align}
\mathcal{L_{\textrm{Yuk}_{\textrm{kink}}}}&\sim
f(\overline{Q_1^{c}}Q_1
-\overline{L^{c}_1}L_1+\overline{Q_2'^{c}}Q_2'-\overline{L^{c}_2}
L_2)\Sigma_1\nonumber\\
&+f'(\overline{Q'^{c}_1}Q'_1 -\overline{L'^{c}_1}L'_1+\overline{Q^{c}_2}Q_2-\overline{L'^{c}_2}L'_2)\Sigma_1\nonumber\\
&+g(-\overline{Q_1^{c}}Q_1 +\overline{L^{c}_1}L_1+\overline{Q_2'^{c}}Q_2'-\overline{L^{c}_2}L_2)\Sigma_2\nonumber\\
&+g'(-\overline{Q'^{c}_1}Q'_1 +\overline{L'^{c}_1}L'_1+\overline{Q^{c}_2} Q_2-\overline{L'^{c}_2}L'_2)\Sigma_2.
\end{align}
If we take $f,f',g,g'>0$, all the right handed fermions are localized
at the ends
of the extra dimension, with quarks at one end and leptons at
the other. Further, we find that $u_R$ ($d_R$) localized about $y=0$
has the same profile as $e_R$ ($\nu_R$) around $y=L$. Meanwhile the
quark and lepton doublets have unrelated profiles with peaks in the
bulk.  This is precisely the setup advocated in \cite{Hung:2002qp,Hung:2004ac} to
achieve a naturally small neutrino Dirac mass. That the leptons are
lighter than
the quarks now results from the lepton doublet been more strongly localised
than the quark doublet. It then
follows that we expect $m_\nu/m_\tau\ll m_b/m_t$ since the difference
in the amplitudes of the right handed wave functions becomes more
dramatic the further in to the bulk we move.

\begin{figure*}[bt]
\centering
\includegraphics[width=0.9\textwidth]{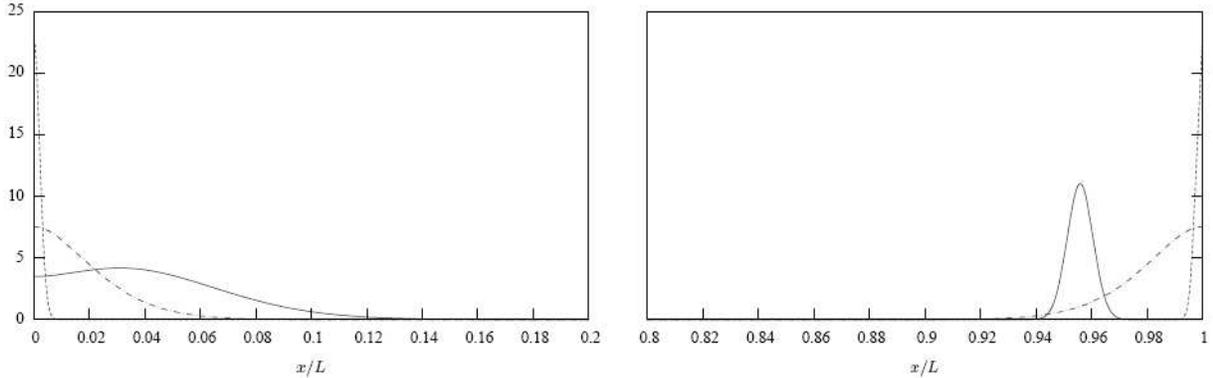}
\caption{Fermion profiles for the parameter values described in Section \ref{sec:sfpdecayandmasses}. Note only the regions around the ends of the extra dimension are plotted as all fermions have miniscule amplitudes in the central region. Quarks (leptons) are shown in the left (right) plot. Solid line=$Q_L (L_L)$, short dash=$d_R (\nu_R)$, dot-dash=$u_R (e_R)$.}
  \label{fig:splitfermions_pdecay}
\end{figure*}

Again simplifying to the one generation case, the parameter choice
$v_1=7.9/(\xi_1 L), v_2=69/(\xi_2 L)$ and
$f=15.6\xi_1,f'=865\xi_1,g=0.440\xi_2,g'=33.3\xi_2$, produces the
fermion localization pattern shown in Figure \ref{fig:splitfermions_pdecay}. Note that the
overlap between quarks and leptons is small enough to
suppress the
proton decay rate below current bounds with an
order 10-100~TeV fundamental scale. If we take $\lambda_t=1.32$,
$\lambda_b=\lambda_\tau=0.0713$ and $\lambda_\nu=0.3$, the fermion
masses are $m_t=173\GeV$, $m_b=4.13\GeV$, $m_\tau=1.78\GeV$ and
$m_\nu=77\meV$.  This setup does contain some hierarchy: for
$v_1\approx v_2$ one requires
$\Lambda_2\approx 100\Lambda_1$ which leads to a
hierarchy of $\mathcal{O}(10^2)$ between the smallest and largest
$\Sigma$ Yukawa coupling. This remains a vast improvement over the
$\sim 12$ orders of magnitude parameter hierarchy required to
explain fermion masses with Dirac neutrinos in the SM. It is
also, to our knowledge, the first realization of the ideas of AS which
implements both features of their proposal. Further work
is required to check that this carries over to three generations and
that, in particular, the SM mixing angles may be reproduced \cite{Twisting}.
However if this
is shown to be the case this would represent the first dynamical setup
to produce both realistic fermion masses and suppress proton decay via
the split fermion mechanism.

\section{Neutral Currents\label{qllr:neutral_currents}}
\begin{figure}
\centering
\includegraphics[width=0.4\textwidth]{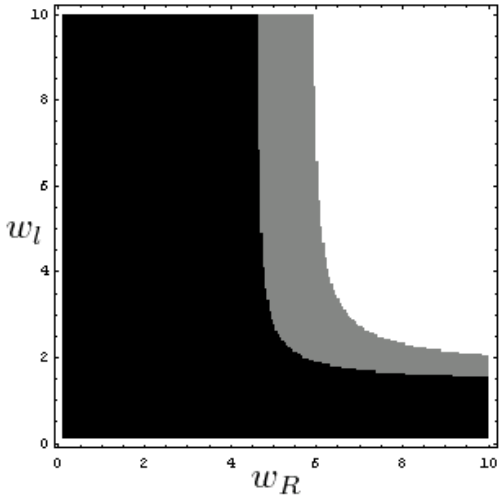}
  \caption{Plot of the values of $w_{R,l}$ which are consistent
    with a $\chi^2$ fit to the electroweak precision parameters (see
    equation (\ref{precision_ew_paramters})\ ).
    The grey (black) region is excluded at the 90\%
    (95\%) confidence level and the units are in TeV.}
  \label{fig:bounds}
\end{figure}

Having specified the fermion content we now present the neutral currents of the model and
obtain bounds on the symmetry breaking scales $w_R$ and
$w_l$. Since we consider $1/R\gg w_l,w_R$ it shall suffice to
consider the interactions of the zero
mode fermions and gauge bosons. After changing to the neutral gauge boson mass eigenstate basis by
diagonalizing the matrix $H$ with the rotation (\ref{eqn:EulerMixing})
the neutral current interactions for the zero mode fields may be written as
\begin{widetext}
\begin{equation}\label{eqn:NeutralCurrent}
{\cal L}_{NC} = \left[ e A^\mu Q ~+~ \left(\frac{g}{
\cos\theta}\right)~ Z_{phy}^\mu~ A_{NC}~+~ \left(g\over
\cos\theta\right)~ Z_{phy}^{\prime\mu}~ B_{NC} ~+~ \left(\frac{g}{
\cos\theta}\right)~ Z_{phy}^{\prime\prime\mu}~ C_{NC}\right]~
J_{NC,~\mu}^{00},
\end{equation}
where the zero mode components of the current are
\begin{equation}
J^{00}_{NC,~\mu}=\sum_{i=1,2}
\bar{Q}^0_i\gamma_{\mu}Q_i^0+\bar{Q}'^0_i\gamma_{\mu}Q_i'^0+\bar{L}^0_i\gamma_{\mu}L^0_i+\bar{L}'^0_i\gamma_{\mu}L'^0_i,
\end{equation}
and we have defined
\begin{eqnarray}
A_{NC} &=& I_{3L}(U_{11}+\delta_Z)-Q(U_{11}\sin^2\theta+\delta_Z)+
I_{3R}(U_{21}\cos\alpha\cos\theta+\delta_Z)+T_{l}U_{31}\frac{g_s\cos\theta}{2
g\cos\beta};\\
B_{NC} &=&
I_{3L}(U_{12}+\delta_{Z'})-Q(U_{12}\sin^2\theta+\delta_{Z'})+
I_{3R}(U_{22}\cos\alpha\cos\theta+\delta_{Z'})+T_{l}U_{32}\frac{g_s\cos\theta}{2
g\cos\beta};\\
C_{NC} &=&
I_{3L}(U_{13}+\delta_{Z''})-Q(U_{13}\sin^2\theta+\delta_{Z''})+
I_{3R}(U_{23}\cos\alpha\cos\theta+\delta_{Z''})+T_{l}U_{33}\frac{g_s\cos\theta}{2
g\cos\beta},
\end{eqnarray}
\end{widetext}
where
\begin{eqnarray}
\delta_{Z}&=&\sin\theta(U_{21}\tan\alpha+U_{31}\sec\alpha\tan\beta),\\
\delta_{Z'}&=&\sin\theta(U_{22}\tan\alpha+U_{32}\sec\alpha\tan\beta),\\
\delta_{Z''}&=&\sin\theta(U_{23}\tan\alpha+U_{33}\sec\alpha\tan\beta),
\end{eqnarray}
where the form of the elements of $U$ may be found in
Appendix~\ref{appendix_a}.
These couplings may now be used to bound the symmetry breaking scales
$w_{R,l}$. We achieve this by performing a $\chi^2$ fit of the
predictions of this model to the following electroweak precision
data:
\begin{eqnarray}
& &R_{e,\mu ,\tau ,b,c}\quad,\quad A_{e,\mu ,\tau ,b,c,s}\quad ,\quad A^{FB}_{e,\mu ,\tau
  ,b,c,s}\quad,\nonumber\\ 
& &Q_W(Cs)\quad ,\quad Q_W(Tl)\quad ,\quad g_{n_L}^2\quad ,\quad g_{n_R}^2\quad,\nonumber\\
& &\Gamma_{Z}\quad,\quad \sigma_{\textrm{had}}\quad.\label{precision_ew_paramters}
\end{eqnarray}
Under the
phenomenological necessary assumption that $M_{Z'}\gg M_Z$,
the physical consequences of the corrections to the coupling of $Z_{phy}$ far
outweigh the new physics resulting from the couplings to $Z'_{phy}$. Thus we
include only this dominant effect when determining our bounds.  We
find that in the LR limit one requires
$w_R>4.8\TeV$ ($w_R>6.2\TeV$) at the 95\% (90\%) confidence
level, which leads to $M_{Z'}>2.7\TeV$ ($M_{Z'}>3.4\TeV$). Note that
we fit to more precision electroweak parameters than
previous works using neutral currents to bound LR breaking scales and
thus, to the best of our knowledge, this is the strongest
bound on $w_R$ from the neutral sector yet obtained in the literature (for previous bounds see
\cite{Czakon:1999ga,Chay:1998hd,Mimura:2002te}). In the QL
limit, it is necessary that $w_l>1.5\TeV$ ($w_l>2.0\TeV$) and
$M_{Z'}>1.6\TeV$ ($M_{Z'}>2.0\TeV$).  If both $w_R$ and $w_l$ are
close to their lower bounds, it is possible that $M_{Z''}$ is also at
the TeV scale.  In this case, at 95\% confidence, we find
$w_R>6.3\TeV, w_l>2.1\TeV$ which leads to $M_{Z'}>2.1\TeV,
M_{Z''}>3.5\TeV$.  These bounds are shown in Figure~\ref{fig:bounds}.
\section{Experimental Signatures\label{qllr:exp}}
In the limit $1/R\gg w_{l,R}$ the key signatures of 5d QLLR models result from the additional neutral gauge bosons and the liptons. The discovery of an additional neutral gauge boson at the LHC would be via Drell-Yann
processes $pp\rightarrow Z'\rightarrow l^+l^-$ \cite{Cvetic:1995zs,Dittmar:2003ir}. The lower center-of-mass energy of a next generation linear collider (NLC) precludes the production of real $Z'$s. Thus any discovery at the NLC would be made by measuring $e^+e^-\rightarrow f\bar f$ and observing corrections resulting from interference between diagrams with a $Z'$ propagator, and those with a $\gamma$ or $Z$ \cite{Cvetic:1995zs}, to the relevant quantities discussed in Section \ref{qllr:neutral_currents}.

Current data allows a $Z'$ at TeV energies regardless of the hierarchy between $w_l$ and $w_R$, however the discovery prospects depend strongly on this hierarchy. In the limit $w_R< w_l$ the additional boson is basically that of the LR model.  Previous studies have found that $Z'_{LR}$ is discoverable up to a mass of about $5\TeV$ at the LHC (or NLC with $\sqrt s=1\TeV$) with an integrated luminosity of $100\mathrm{fb}^{-1}$ \cite{Cvetic:1995zs,Rizzo:1996ce}. If $w_l< w_R$ the light $Z'$ is that of the QL model.  In this case discovery at a hadron collider such as the LHC is unlikely since the cross section of the Drell-Yann process, $\sigma(pp\rightarrow Z')B(Z'\rightarrow l^+l^-)$, will not be large. This is since $Z'$ contains a large fraction of $G^0_l$ which does not couple to the quarks, meaning the cross section $\sigma(pp\rightarrow Z')$ will be small.  The discovery prospects are greater at the NLC since $G^0_l$ has large universal couplings to all leptons.  These couplings are quite different from any of the canonical $Z'$s usually considered, providing a distinctive signature.

The most interesting case is when $w_R\sim w_l$. Then both $Z'$ and $Z''$ possess significant couplings to the $SU(3)_l$ and $SU(2)_R$ neutral currents and may have TeV masses.  As such they could both be produced at the LHC.  The prospects of discovery depend on the masses of the liptons, which are of order $w_l$.  If the $Z'$ bosons can decay into liptons their branching fraction into charged leptons will reduce and hence the Drell-Yann cross section will be smaller. The presence of light liptons would afford alternative discovery channels, with liptonic final states, at the LHC.  At a linear collider light liptons are unimportant since only virtual $Z'$s are produced.  However, diagrams with both $Z'$ and $Z''$ propagators will interfere with those of the SM, causing corrections.  Without prior knowledge of the mass and mixings of the neutral gauge bosons their couplings to fermions are unknown. Thus separating their effects and categorically identifying the nature of the bosons is difficult.

If $w_l$ is at the TeV scale then the production of liptons is also possible at the LHC. For $1/R\gg w_l$, only the zero mode liptons with orbifold parity $(+,+)$, which are in different $SU(3)_l$ multiplets to the SM leptons, can be produced.  Since both the $Y_{1,2}^\pm$ and $W_R$ bosons have $(+,-)$ orbifold parity, they do not directly couple the lightest liptons to the SM leptons.  As the unbroken $SU(2)_l$ is expected to be confining these liptons will form bi-lipton bound states. These states will decay into SM fermions via creation of a $W$, $Z$ or photon.  This will produce a clear experimental signature, the details of which are similar to those obtained in the 4d QLLR model~\cite{Foot:1991fk}.
 
\section{Conclusion}\label{qllr_conc}
In this paper we have constructed and analysed the 5d QLLR model. We
have shown that the higher dimensional construct permits a novel
mechanism for suppressing neutrino masses below the electroweak scale and
allows one to dramatically simplify the scalar sector employed in
4d constructs. This allows one to keep both the QL and the LR symmetry
breaking scales low (TeV energies) so that two neutral gauge bosons
may be observed at the LHC.

Split fermions were used to explain
some of the features of the SM mass spectrum. Two interesting
fermionic geographies were presented, each of which provided a
rationale for the relationships $m_t>m_b,
m_{\tau}$ and $m_\nu\ll m_t$. One of these had no Yukawa coupling
hierarchy but to suppress the proton decay
rate required either a large cut-off or that the fundamental theory observed the accidental $B$ and $L$ symmetries of the QLLR model. In the former case, fine tuning was required to stabilize the Higgs boson
mass at the electroweak scale. The alternative arrangement
suppressed the proton
decay rate by spatially separating quarks and leptons in the extra
dimension. Thus the
hierarchy problem was
alleviated but at the cost of introducing an order $10^2$ Yukawa
coupling hierarchy. Given the extent to which the symmetries of the
model constrain the Yukawa sector it is a non-trivial result that
interesting fermionic geographies can be obtained with mild Yukawa
coupling hierarchies. These arrangements show promise but further work
is required to ensure that a fully realistic three generational setup
may be obtained.
\section*{Acknowledgements}
The authors thank R. Foot, G. Joshi, B. McKellar and R. Volkas. This work was
supported in part by the Australian Research Council.
\appendix
\section{Neutral Gauge Boson Mixing}\label{appendix_a}
The physical $Z$-bosons are found by performing a 3-dimensional
orthogonal rotation of
the interaction $Z$-bosons. Defining
\begin{eqnarray}
Z_P&=&(Z_{phy},Z_{phy}',Z_{phy}'')^T,\nonumber\\
Z_I&=& (Z,Z',Z'')^T,
\end{eqnarray}
the relationship between the physical and interaction states is
\begin{eqnarray}\label{eqn:urotation}
Z_P=U^{-1}Z_I,
\end{eqnarray}
with the diagonalized
mass matrix being $D=U^{-1} H U$ and $H$ is defined in (\ref{eqn:ZMixing}).
We will parameterize the rotation matrix as
\begin{widetext}
\begin{eqnarray}\label{eqn:EulerMixing}
U(\zeta,\sigma,\psi)\equiv\left(
\begin{array}{ccc}
 \cos \zeta \cos \psi  -
    \cos \sigma \sin \zeta \sin \psi &
   \cos \psi \sin \zeta +
    \cos \zeta\cos \sigma\sin \psi &
   \sin \sigma \sin \psi  \\
   -\cos \sigma \cos \psi \sin \zeta  -
    \cos \zeta \sin \psi &
   \cos \zeta \cos \sigma \cos \psi  -
    \sin \zeta \sin \psi & \cos \psi \sin \sigma  \\
   \sin \zeta \sin \sigma &
   - \cos \zeta \sin \sigma  & \cos \sigma
   \end{array}\right),
   \end{eqnarray}
\end{widetext}
and we note that $U^{-1}(\zeta,\sigma,\psi)=U(-\psi,-\sigma,-\zeta)$.
To order $\mathcal{O}\left(\frac{k^4}{w^4}\right)$ we find
\begin{eqnarray}\label{eqn:MixingRelations}
U_{11}=1,
\end{eqnarray}
\begin{eqnarray}
U_{21}=
\frac{m_Z^2\cos^3\alpha\cos\theta(9w_R^2\sin^4\beta+4w_l^2)}{2g^2w_l^2w_R^2},
\end{eqnarray}
\begin{eqnarray}
U_{31}=\frac{-3\sqrt{3}m_Z^2\cos^2\alpha\cos\beta\cos\theta\sin^2\beta}{2
gg_sw_l^2},
\end{eqnarray}
\begin{eqnarray}\label{eqn:U33}
U_{33}=U_{22}=\sqrt{\mu_{+}}-m_Z^2\frac{g^2w_R^2}{4\Delta^2}\sin
2\theta \tan\beta \sqrt{\mu_{-}},
\end{eqnarray}
\begin{eqnarray}
U_{23}=-U_{32}=\sqrt{\mu_{-}}+m_Z^2\frac{g^2w_R^2}{4\Delta^2}\sin
2\theta \tan\beta \sqrt{\mu_{+}},
\end{eqnarray}
\begin{eqnarray}
U_{12}&=&-
\left(\frac{m_Z^2}{M^2-\frac{1}{2}\Delta}\right)U_{33}\cos\alpha\cos\theta,\\
U_{13}&=&
\left(\frac{m_Z^2}{M^2+\frac{1}{2}\Delta}\right)U_{32}\cos\alpha\cos\theta.
\end{eqnarray}
Expressing the mixing angles in terms of the elements $U_{ij}$ we have
 \begin{eqnarray}
\sigma &=&\arccos[U_{33}],\\
\zeta &=&\arcsin[U_{31}\csc\sigma],\label{eqn:zeta}\\
\psi &=&\arcsin[U_{13}\csc\sigma],\label{eqn:psi}
\end{eqnarray}
if $\sigma \neq 0$ and 
\begin{equation}
\psi+\zeta =-\arcsin U_{21}\label{eqn:PsiEqualsZeta},
\end{equation}
if $\sigma=0$.
\section{Large LR Breaking Limit}\label{large_w_l_limit_appendix}
In the limit $w_R\rightarrow\infty$ one has $U_{13}\rightarrow 0$ so that
$\psi\rightarrow0$. Thus
\begin{eqnarray}\label{eqn:Uql}
U=\left(\begin{array}{ccc}
\cos\zeta&\sin\zeta&0\\-\cos\sigma\sin\zeta&\cos\zeta\cos\sigma&\sin\sigma\\
\sin\zeta\sin\sigma &-\cos\zeta\sin\sigma &\cos\sigma
\end{array}\right)
\end{eqnarray}
This may be written as $U=R_{\sigma}R_{\zeta}$ where
\begin{eqnarray}\label{eqn:R_sigma}
R_\sigma=\left(\begin{array}{ccc}
1&0&0\\0&\cos\sigma&\sin\sigma\\0&-\sin\sigma&\cos\sigma
\end{array}\right),
\end{eqnarray}
\begin{eqnarray}\label{eqn:R_zeta}
R_\zeta=\left(\begin{array}{ccc}
\cos\zeta&\sin\zeta&0\\-\sin\zeta&\cos\zeta&0\\
0&0 &1
\end{array}\right),
\end{eqnarray}
so that
\begin{eqnarray}\label{eqn:urotation_large_wr}
Z_P&=&U^{-1}Z_I\nonumber\\
&=&R_\zeta^{-1}R_\sigma^{-1}Z_I\nonumber\\
&=&R_\zeta^{-1}\tilde{Z}_I,
\end{eqnarray}
where we have redefined the interaction basis as
$\tilde{Z}_I=R_\sigma^{-1}Z_I$.
In this basis, one of the additional neutral gauge bosons, $Z''_{QL}$, decouples in the large $w_R$ limit. The mixing between the SM $Z$ boson and the other additional neutral gauge boson, $Z'_{QL}$, in this limit 
is given by $\zeta$. These extra gauge boson
are 
\begin{eqnarray}
Z_{QL}'=\cos\sigma Z' -\sin\sigma Z'',\\
Z_{QL}''=\sin\sigma Z' +\cos\sigma Z'',
\end{eqnarray}
with $Z'$ and $Z''$ defined in (\ref{eqn:zdoubleprime}) and the subscript emphasises that $Z_{QL}'$ is the extra neutral boson found in the QL symmetric model. 



\begin{thebibliography}{999}
\bibitem{mohapatra_error} There
is a disagreement about the power of the $\cos^4\alpha$ term as \cite{Mohapatra:2002rn} has a power of two. We have
corresponded with the authors and they confirm a misprint.
\bibitem{BLsymm}
Such an approach may be taken consistently with the requirement for negligible Majorana neutrino masses.  There are values for $w_l,w_R,\Lambda$ such that the Majorana masses will be sufficiently small to ignore but proton decay will proceed at rates above experimental bounds.  Alternatively, the  theory above the cut-off may preserve only $B-L$, allowing proton decay to proceed but forbidding Majorana neutrino masses.
\bibitem{running_masses} The masses must be run to a common scale
  which we take to be $m_t$.  This leads to running constants
  $m_b(m_b)/m_b(m_t)=1.55,\quad m_\tau(m_\tau)/m_\tau(m_t)=1.02$.
\bibitem{Twisting}
One complication when we begin considering a three generation model is that the fermion eigenstates which couple to $\Sigma_1$ need not coincide with the eigenstates coupling to $\Sigma_2$.  Most previous split fermion work, such as that listed in \cite{SFFlavour}, has ignored this complication and assumed the coupling matrices are simultaneously diagonalizable.  The implications when such `twisting' is considered are discussed in \cite{Grossman:2004rm}.
\end{thebibliography}
\end{document}